\documentclass[twocolumn]{aastex62}
\usepackage{amsmath}
\usepackage{natbib}
\pdfoutput=1

\shortauthors{Gustafsson et al.}

\newcommand{\spitzer}{{\it Spitzer}} 
\newcommand{\farcsec}{\hbox{$.\!\!^{\prime\prime}$}}
\newcommand{\farcmin}{\hbox{$.\!\!^{\prime}$}}

\begin{document}

\title{Spitzer Albedos of Near-Earth Objects}

\correspondingauthor{Annika Gustafsson}
\email{ag765@nau.edu, agustafsson@lowell.edu}

\author{Annika Gustafsson}
\affil{Department of Physics and Astronomy, 
P.O. Box 6010, 
Northern Arizona University,
Flagstaff, AZ 86011}

\author{David E. Trilling}
\affil{Department of Physics and Astronomy, 
P.O. Box 6010, 
Northern Arizona University,
Flagstaff, AZ 86011}

\author{Michael Mommert}
\affil{Lowell Observatory,
1400 W Mars Hill Rd.,
Flagstaff, AZ 86011}

\author{Andrew McNeill}
\affil{Department of Physics and Astronomy, 
P.O. Box 6010, 
Northern Arizona University,
Flagstaff, AZ 86011}

\author{Joseph L. Hora}
\affil{Center for Astrophysics \textbar Harvard \& Smithsonian,
60 Garden Street, MS-65,
Cambridge, MA 02138}

\author{Howard A. Smith}
\affil{Center for Astrophysics \textbar Harvard \& Smithsonian,
60 Garden Street, MS-65,
Cambridge, MA 02138}

\author{Stephan Hellmich}
\affil{Institute of Planetary Research, 
Deutsches Zentrum f\"{u}r Luft-
und Raumfahrt e.\ V. (DLR), 
Rutherfordstrasse 2, 12489 Berlin, Germany}

\author{Stefano Mottola}
\affil{Institute of Planetary Research, 
Deutsches Zentrum f\"{u}r Luft-
und Raumfahrt e.\ V. (DLR), 
Rutherfordstrasse 2, 12489 Berlin, Germany}

\author{Alan W. Harris}
\affil{Institute of Planetary Research, 
Deutsches Zentrum f\"{u}r Luft-
und Raumfahrt e.\ V. (DLR), 
Rutherfordstrasse 2, 12489 Berlin, Germany}

%%%%%%%%%%%%%%%%%%%%%%%%%%%%%%%%%%%
%%%%%		   		 ABSTRACT 				%%%%%
%%%%%%%%%%%%%%%%%%%%%%%%%%%%%%%%%%%
\begin{abstract}

Thermal infrared observations are the most effective way to measure asteroid diameter and albedo for a large number of near-Earth objects. Major surveys like NEOWISE, NEOSurvey, ExploreNEOs, and NEOLegacy find a small fraction of high albedo objects that do not have clear analogs in the current meteorite population. About 8\% of \spitzer-observed near-Earth objects have nominal albedo solutions greater than 0.5. This may be a result of lightcurve variability leading to an incorrect estimate of diameter or inaccurate absolute visual magnitudes. For a sample of 23 high albedo NEOs we do not find that their shapes are significantly different from the \cite{2019McNeill} near-Earth object shape distribution. We performed a Monte Carlo analysis on 1505 near-Earth objects observed by \spitzer, sampling the visible and thermal fluxes of all targets to determine the likelihood of obtaining a high albedo erroneously. Implementing the McNeill shape distribution for near-Earth objects, we provide an upper-limit on the geometric albedo of $0.5\pm0.1$ for the near-Earth population.

\end{abstract}


%%%%%%%%%%%%%%%%%%%%%%%%%%%%%%%%%%%
%%%%%			 INTRODUCTION 				%%%%%%
%%%%%%%%%%%%%%%%%%%%%%%%%%%%%%%%%%%

\section{Introduction} \label{sec:intro}

Asteroids are some of the most primitive bodies in our Solar System and hold important clues regarding Solar System formation and evolution. Near-Earth objects (NEOs) are the subset of small bodies whose orbits bring them close to the Earth, with perihelion distances less than 1.3~au. They are a diverse population that represent the composition and dynamics throughout the Solar System. Due to their close approach to Earth, they pose a threat as potential impactors, present a unique opportunity as mission accessible targets, and are a potential source of mining resources. As a result, it is important we understand the physical properties of these objects. 

Only absolute visual magnitudes (H-magnitudes) and orbital parameters are known for most NEOs. On discovery, the only physical information available for an NEO is its visible brightness, and the current rate of discovery of NEOs is far exceeding progress in physical characterization (e.g., albedo, diameter, composition). As a result, physical characterization surveys are essential to help inform hazard assessment, mitigation efforts, and mission planning for nearby NEOs. These surveys also provide an improved understanding of the distribution of physical properties which can inform NEO population models \citep{2018Granvik}.  

Thermal infrared observations are critical for determining diameter and albedo for a large number of targets. ExploreNEOs \citep{2010Trilling}, IRAS \citep{2010Ryan}, AKARI \citep{2014bOkamura, 2014Okamura}, NEOSurvey \citep{2016Trilling}, NEOLegacy, and NEOWISE \citep{2016Nugent, 2017Masiero, 2018Masiero} are infrared characterization surveys that determine diameter and albedo of small bodies across the Solar System. Albedos are particularly important in helping to infer composition when ambiguities exist in the classification of visible and near-infrared reflectance spectra \citep{2003Delbo}. Albedo and diameter are derived using thermal models \citep{1998Harris, 2003Delbo, 2006Mueller, 2007Harris, 2015Delbo}, which assume equilibrium between absorbed solar energy and thermal emission \citep{2003Delbo}. 

\begin{figure}
  \includegraphics[width=3.4in]{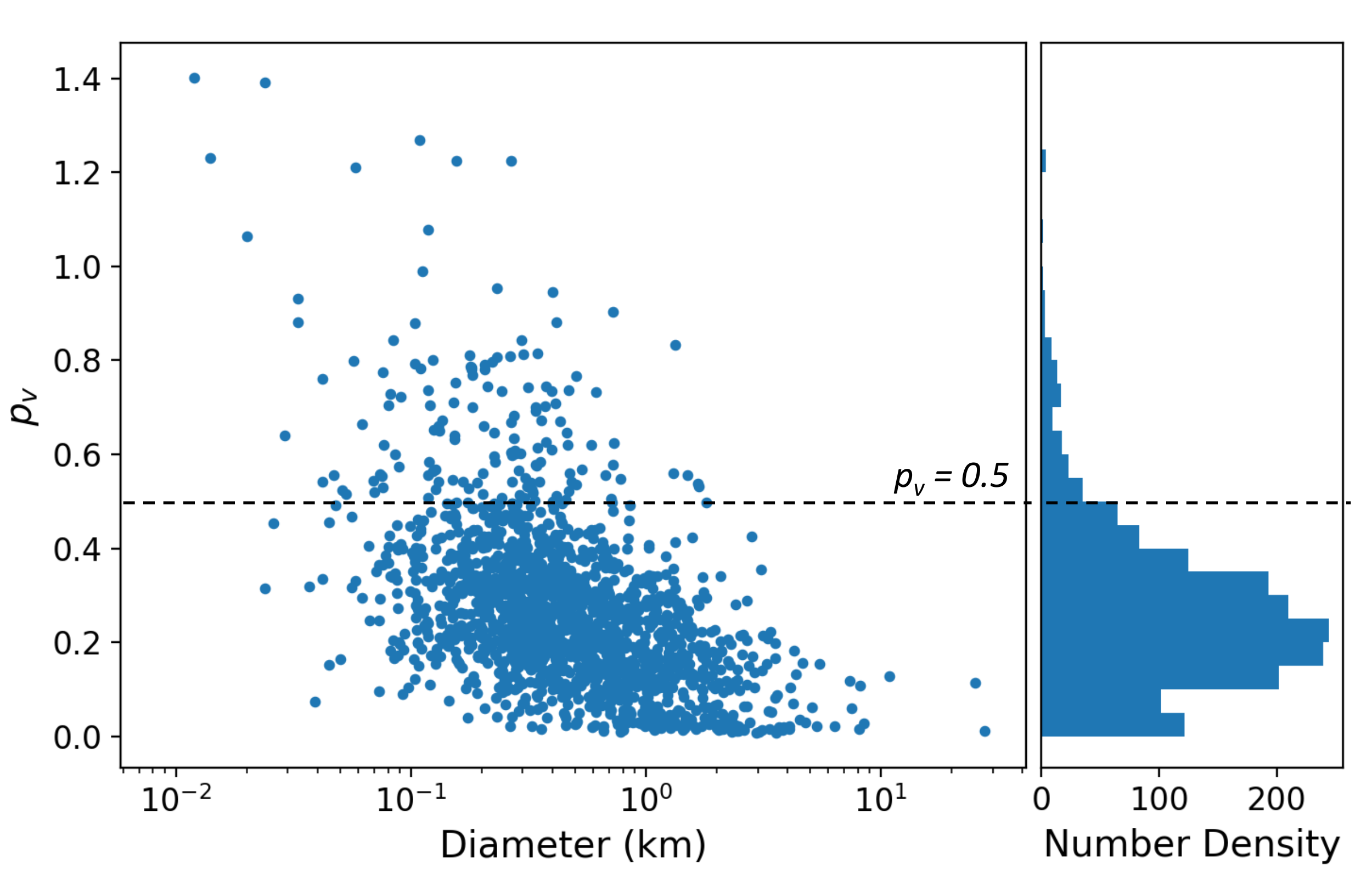}
  \caption{Diameter-albedo distribution of targets. The left panel is the diameter-albedo relationship for the \spitzer~sample. The right panel is the number density as a function of albedo. $\sim$8\% of the targets in this sample have albedo values greater than 0.5, a factor of about 8 more than expected for the NEO population from work by \cite{2011Thomas} and \cite{2004Stuart}. \label {fig:Sp_pv}}
\end{figure}

ExploreNEOs, NEOSurvey, and NEOLegacy observed over 1500 near-Earth objects as of June 2017 using the \spitzer/IRAC instrument \citep{2004Fazio}. 121 targets in the 2017 sample ($\sim8\%$) have nominal derived albedos ($p_V$) greater than 0.5 (Figure~\ref{fig:Sp_pv}). Using the \cite{2011Thomas} average albedo for taxonomic complexes in combination with the \cite{2019Binzel} NEO taxonomy distribution, we expect only $1.04 \pm 0.09\%$ of a sample ($\sim21$ targets in this case) to have albedo values greater than or equal to 0.5. We find \spitzer~high albedo targets in excess by a factor of about 8. As a result, \cite{2010Trilling, 2016Trilling} suggest that the most likely explanations for such high albedos are H-magnitudes that are photometrically inaccurate and/or sampling bias of objects from high amplitude light curves that give a high albedo solution.   

%%%%%%%%%%%%%%%%%%%%%%%%%%%%%%%%%%%
%%%%%		LIGHTCURVE EFFECTS ON ALBEDO	%%%%%%
%%%%%%%%%%%%%%%%%%%%%%%%%%%%%%%%%%%

\section{Lightcurve Effects on Albedo} 

NEOs with large lightcurve amplitudes can lead to high albedo values. The albedo-amplitude relationship is derived from the diameter-albedo equation \citep{1989Bowell}: 
\begin{equation} \label{eq:pv}
	D = \frac{1329}{\sqrt{p_V}} 10^{-0.2 H}
\end{equation}

For a poorly sampled lightcurve, the lightcurve amplitude can have a strong effect on the derived albedo. We expect the lightcurve induced albedo discrepancies are strong in our \spitzer~observations as we only observe targets for a relatively short period of time. Using Equation~\ref{eq:pv} to solve for $p_V$, we find that the derived albedo falls in the range shown in Equation~\ref{eq:pv2} below where $A$ is the peak to trough lightcurve amplitude. 
\begin{equation} \label{eq:pv2}
	[\frac{1329}{D} 10^{-0.2(H-\frac{A}{2})}]^2 \le p_V \le [\frac{1329}{D} 10^{-0.2(H+\frac{A}{2})}]^2
\end{equation}

\begin{figure*}
  \includegraphics[width=\textwidth]{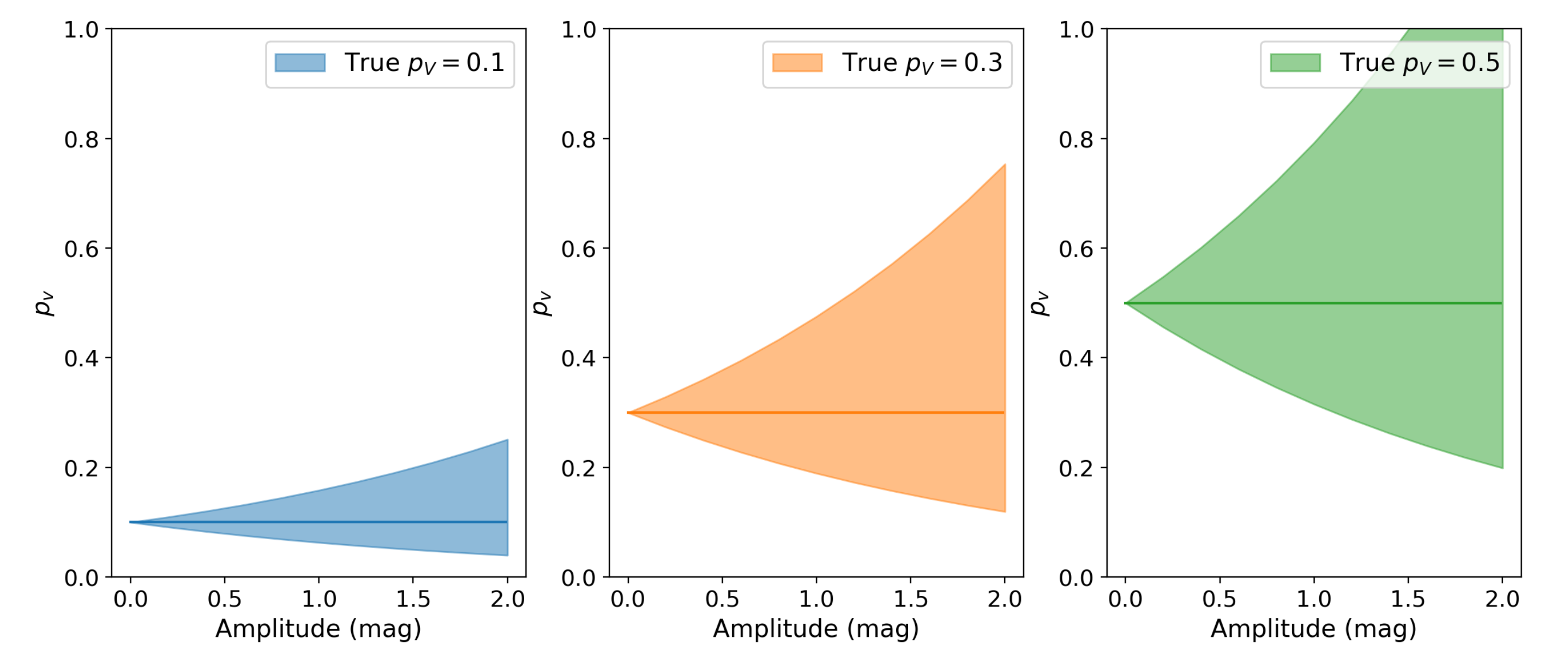}
  \caption{Amplitude-albedo distribution for a generic NEO ($H=19$). By poorly sampling the lightcurve for an elongated object, and thus, miscalculating H, the albedo of the target can easily be determined to be greater than or equal to 0.5. Each panel indicates an NEO whose true albedo is shown for a lightcurve amplitude of 0, but can have an identified albedo anywhere in the shaded region for varying lightcurve amplitudes if the H-magnitude is incorrectly identified. For larger intrinsic true albedo values, a smaller lightcurve amplitude is required to allow for albedo values greater than 0.5. \label {fig:pv_amp}}
\end{figure*}

Figure~\ref{fig:pv_amp} shows the albedo-amplitude relationship for a generic NEO ($H=19$) with different visible lightcurve amplitudes from 0 to 2~magnitudes and different characteristic albedo values. The shaded regions in Figure~\ref{fig:pv_amp} represent the distribution of albedo misestimates for each lightcurve amplitude. The solid line is the mean albedo derived for each amplitude and is always the true albedo of the target. For a target whose true albedo value is 0.3, a 1~magnitude lightcurve amplitude can easily bring the derived albedo above 0.5. For a target whose true albedo is larger, a derived albedo greater than 0.5 can be achieved at much smaller lightcurve amplitudes. 

To test whether the observed high albedos in the data represent truly high albedos or modeling outliers, we employ a Monte Carlo analysis of \spitzer~NEOs to investigate the contributions of poorly sampled lightcurves and large thermal model input uncertainties. Table~\ref{tab:definitions} outlines terminology, data samples, and populations that will be used in this work. In Section~\ref{sec:obs} we investigate high amplitude lightcurves with photometry. We outline the methods and results of our Monte Carlo technique in Sections~\ref{sec:MC1} and~\ref{sec:results} and discuss possible true high albedo targets in Section~\ref{sec:HA_NEOs}. 

%%%%

\begin{deluxetable*}{ccc} 
\tablecaption{Glossary of Import Terms and Data Samples \label{tab:definitions}}
\tablecolumns{2}
\tablenum{1}
\tablewidth{0pt}
\tablehead{
\colhead{Name}	& & \colhead{Definition}
}	
\startdata
{\it{Terminology:}}	&&		\\
High Albedo				&		& Albedo greater than or equal to 0.5 \\
Intrinsic Albedo 			&		& Inherent albedo of the target \\
Nominal Albedo 			&		& Derived albedo of the target \\
\hline
{\it{Spitzer Data Samples:}} &{\it{\#}}&			\\
\spitzer~2017				& 1505 	&NEOs observed with ExploreNEOs, NEOSurvey, \\ 
						&		&			and NEOLegacy as of JD 2457921 \\
\spitzer~2017 High Albedo 	& 121 	&NEOs in the \spitzer~2017 sample with albedo\\ 							&		& 		greater than or equal to 0.5 \\
DCT Photometry			& 22 		&\spitzer~2017 High Albedo NEOs observed with DCT\\
Calar Alto Photometry 		& 7 		& 6 of the 22 DCT High Albedo NEOs \& 1 other High Albedo \\ 
						&		&	NEO observed at Calar Alto Astronomical Observatory \\ 
LCDB					& 25 	&  \spitzer~2017 High Albedo NEOs with previously measured \\
						&		& lightcurves reported to the Asteroid Lightcurve Database\\
\spitzer~2018 				& 1999 	& NEOs observed with ExploreNEOs, NEOSurvey, \\ 
						&		& and NEOLegacy as of JD 2458278 \\
\spitzer~2018 High Albedo 	& 181 	& NEOs in the \spitzer~2018 sample with albedo \\
						&		&	greater than or equal to 0.5 \\
Measured Taxonomy 		& 54 	& NEOs in the \spitzer~2018 High Albedo sample with \\ 
						&		& a measured Bus-DeMeo taxonomy classification \\
\hline
{\it{Populations:}} &&			\\
Observed Fraction			&		& \spitzer~2017 albedo distribution\\
MC Fraction				&		& Monte Carlo albedo distribution using the \spitzer~2017 data sample\\
Residual					&		& Fractional difference between the Observed fraction and MC fraction\\
Observed NEO Population	&		& Albedo distribution using the mean taxonomic albedos with \\
						&		& 	the NEO taxonomy distribution \\
MC Fraction of Real $p_V$	&		& Also known as MC Fraction \\ 
						& 		& Same as Residual but normalized to 100\% for albedo greater than 0\\
						&		&  [Treated as an independent population] \\
High Albedo Sample			&		& Same as Measured Taxonomy sample described above\\
\enddata
\end{deluxetable*}

%%%%%%%%%%%%%%%%%%%%%%%%%%%%%%%%%%%
%%%%%%%  				OBSERVATIONS 		%%%%%%
%%%%%%%%%%%%%%%%%%%%%%%%%%%%%%%%%%%

\section{Observations} \label{sec:obs}

%%%%%%%%%%%%%%%%%%%%%%%%%%%%%%%%%%%

\subsection{Spitzer Space Telescope}

The target sample considered for this work consists of 1505 NEOs observed with the {\it Spitzer}/IRAC instrument \citep{2004Fazio} as part of ExploreNEOs \citep{2010Trilling}, NEOSurvey \citep{2016Trilling}, and NEOLegacy. This dataset contains all targets observed by these \spitzer~surveys as of JD 2457921 (2017-06-16). We will refer to this dataset as the \spitzer~2017 sample. ExploreNEOs collected observations in the 3.6 and 4.5~$\micron$ bands, while NEOSurvey and NEOLegacy observed targets in only the 4.5~$\micron$ band. For this work, we consider thermal fluxes from the longer wavelength band where the NEO is typically brightest. A table of the observations and results is maintained in an online database\footnote{\url{http://nearearthobjects.nau.edu}}. A detailed description of the observing methods and lightcurve results are provided in \cite{2018Hora}. 

Diameters listed in the online database are obtained with a modified version of NEATM described in \cite{2016Trilling}. This modified NEATM is the \cite{1998Harris} Near-Earth Asteroid Thermal Model (NEATM) to fit diameter and geometric albedo in combination with a Monte Carlo simulation to estimate the uncertainties associated with the output diameter and albedo. The model uses input \spitzer~fluxes from the 4.5~$\micron$ band and absolute H-magnitudes from the Asteroids Dynamic Site\footnote{\url{http://hamilton.dm.unipi.it/astdys/}} (AstDyS) \citep{2012Knezevic}. The \cite{2012Pravec} correction is applied to the AstDyS H-magnitudes to create the input for the NEATM model. The corrected AstDyS H-magnitudes have a mean 1--sigma uncertainty of 0.24 mag. NEATM also uses a variable beaming parameter ($\eta$) to account for surface roughness, thermal inertia, and other effects in a zeroth order approximation \citep{1998Harris}. A linear relation is used to derive an appropriate $\eta$ based on the phase angle ($\alpha$) where larger $\eta$ is used for higher phase angle observations. This relationship between $\eta$ and $\alpha$ is not well constrained and is thus primarily responsible for the overall uncertainties in the model results \citep{2016Trilling}. The uncertainty in $\eta$ derived in this way contributes up to 90\% of the overall uncertainty in albedo for all objects \citep{2016Trilling}.

There are 121 NEOs from the \spitzer~2017 sample with nominal derived albedo values greater than 0.5. We collected lightcurve photometry for 23 of these objects to place limits on the visible lightcurve amplitudes and determine if lightcurve effects could be a significant contributor to the high albedo values found in the \spitzer~dataset. 22 objects were observed using the Lowell Observatory 4.3--m Discovery Channel Telescope (DCT) in Happy Jack, AZ. We observed 6 of those 22 DCT targets along with one other high albedo target using the 1.23--m Calar Alto Astronomical Observatory in Almer\'ia, Spain. 

%%%%%%%%%%%%%%%%%%%%%%%%%%%%%%%%%%%

\subsection{Discovery Channel Telescope}

Data for 22 high albedo NEOs were collected using the Large Monolithic Imager (LMI) on the DCT. LMI has a 12$\farcmin5\times12\farcmin5$ field of view with a $0\farcmin12$/pixel unbinned pixel scale. Images were taken in 2$\times$2 binning mode for a $0\farcsec24$/pixel on seven separate nights from 2015 to 2017. All targets were observed in the Johnson V filter and at sidereal rates with exposure times ranging from 15 to 60 seconds. The net integration time of a given target ranged from 10 minutes to 2 hours with a majority of the targets observed in the 10--20 minute range, resulting generally in partial lightcurves. A total of 36 separate observations were made of the 22 unique targets. A full description of the DCT observations is provided in Table~\ref{tab:dct_obs}. 

All data were bias-corrected and flat-fielded using biases and twilight flats taken at the beginning of each observing night. On the night of 2016--05--30, no twilight flats were taken, so the flat field was created using sky flats where the first images at each pointing were median combined and normalized to create a master flat. 

%%%%%%%%%%%%%%%%%%%%%%%%%%%%%%%%%%%

\subsection{Calar Alto Astronomical Observatory}

Calar Alto data were collected using the DLR-MKIII camera on the 1.23-m telescope. The camera is a 4000$\times$4000 pixel array with a pixel scale of $0\farcsec32$/pixel and a field of view of $21\farcmin5\times21\farcmin5$. 

We observed 7 high albedo targets over multiple nights. A total of 61 separate observations were made of the 7 unique targets. The filters of the targets varied between V, R, and clear within and across different nights of observations to measure colors and assign composition. The targets were observed over varying phase angle ($4.4^{\circ} \le \alpha \le 75.5^{\circ}$). All images were bias-corrected and flat-fielded. A full description of the Calar Alto observations is provided in Table~\ref{tab:caloalto_obs}.

%%%%%%%%%%%%%%%%%%%%%%%%%%%%%%%%%%%

\subsection{Photometry} \label{sec:Phot}

Automated photometric analysis was performed on all of the reduced DCT and Calar Alto images using the Photometry Pipeline \citep{2017Mommert}. Photometry Pipeline uses Source Extractor \citep{1996Bertin} and SCAMP \citep{2006Bertin} to perform aperture photometry and astrometry on all objects in the field of view. The tool utilizes a curve-of-growth analysis to optimize the aperture size. All targets were located using the Pan-STARRS reference catalog \citep{2017Chambers}. We performed visual rejection for exposures within the proximity of a nearby star or where the target was not correctly identified. Visual inspection was favored as it usually leads to a more conservative rejection of potentially compromised photometric measurements as compared to an automated rejection scheme based on Source Extractor flags.

%%%%%%%%%%%%%%%%%%%%%%%%%%%%%%%%%%%
%%%%%%%  					SHAPE	 		%%%%%%
%%%%%%%%%%%%%%%%%%%%%%%%%%%%%%%%%%%

\section{NEO Shape Distribution}

We determined lower limit amplitudes ($\Delta m$) for all 23 targets using Equation~\ref{eq:deltam} from the photometry where $m$ is the magnitude of the target.
\begin{equation} \label{eq:deltam}
\Delta m = m_{max} - m_{min}
\end{equation}
The lower limit amplitudes are listed in Table~\ref{tab:dct_obs} and Table~\ref{tab:caloalto_obs}. We compare these amplitudes to the \cite{2019McNeill} NEO shape distribution. 

We combined the lower limit amplitudes for our 23 DCT and Calar Alto targets with 25 high albedo targets from the Asteroid Lightcurve Database (LCDB) \citep{2009Warner}. These 25 targets have albedo values greater than 0.5 in the \spitzer~2017 dataset and have previously observed lightcurve information (Table~\ref{tab:LCDB}). The lower limit amplitudes of these 48 high albedo targets are phase corrected using the \cite{1990Zappala} phase-amplitudue relationship (Equation~\ref{eq:phase}) which describes a linear relationship for phase angles from $0^{\circ}<\alpha<40^{\circ}$. In Equation~\ref{eq:phase}, $A$ is the peak to trough lightcurve amplitude, $\alpha$ is the phase angle of the observation, and $m$ is the slope parameter defined for each taxonomy. 
\begin{equation} \label{eq:phase}
A(\alpha=0^{\circ}) = \frac{A(\alpha_{obs})}{1+m\alpha}
\end{equation}

\begin{figure} 
  \includegraphics[width=3.4in]{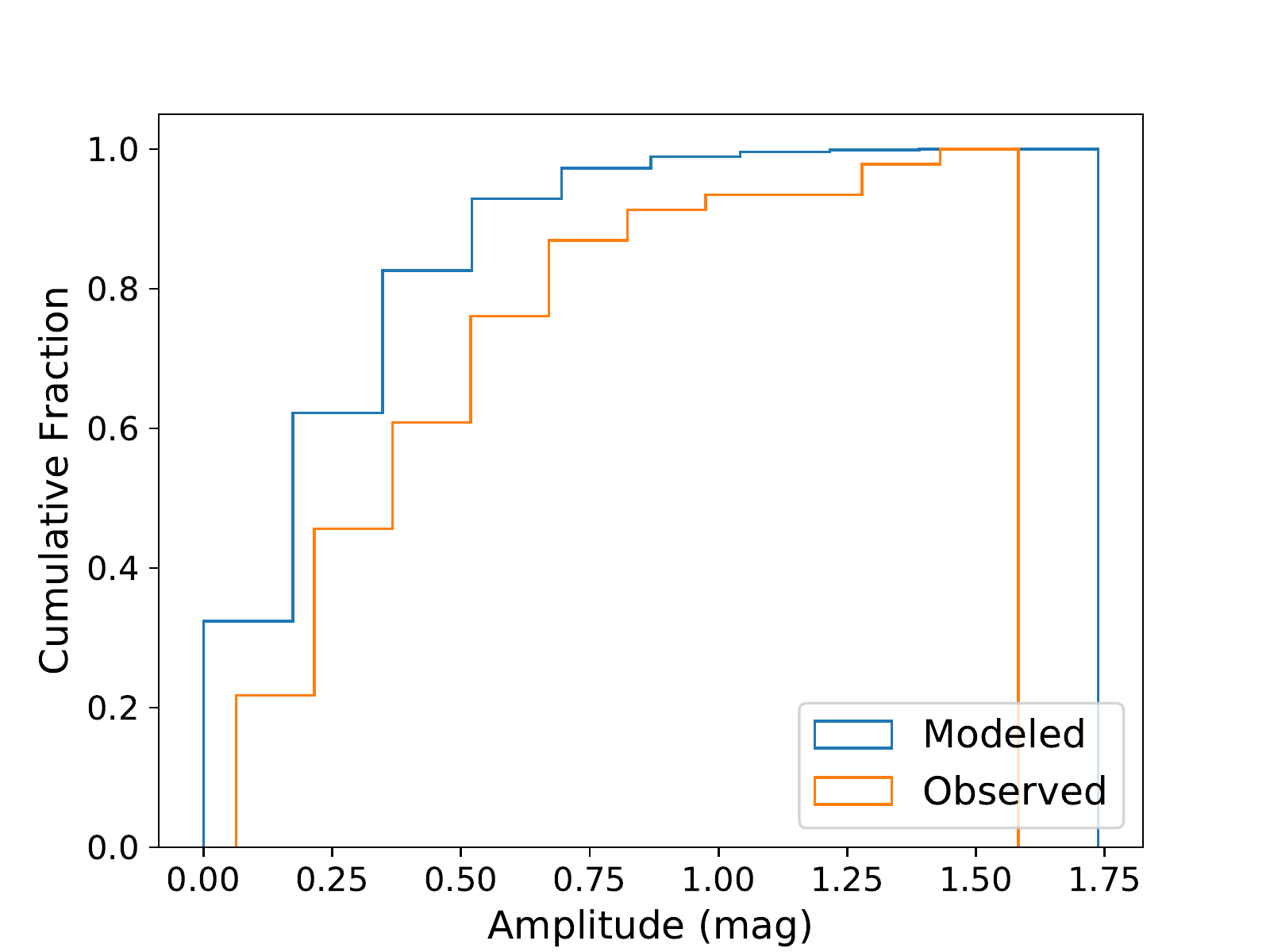}
  \caption{Cumulative NEO amplitude distribution derived from the \cite{2019McNeill} shape distribution (Modeled; blue). Overplotted in orange is the observed cumulative lower limit amplitude distribution of 23 targets from DCT and Calar Alto and 25 targets with previous observations submitted to the LCDB \citep{2009Warner}. All amplitudes have been corrected for phase angle following the relationship derived in \cite{1990Zappala}. We find three candidates for extreme elongated objects in our sample (i.e. objects with a phase-corrected lightcurve amplitude $>1$~magnitude \citep{2018McNeill}) and a maximum lower limit amplitude in our sample of 1.58~magnitudes. \label{fig:DeltaV}}
\end{figure}

The observations of these 48 high albedo targets cover a phase angle range from $3^{\circ}<\alpha<84.5^{\circ}$. Using a slope parameter of $m=0.015$ for all targets and Equation~\ref{eq:phase}, we derive a cumulative distribution of corrected lower limit amplitudes. For phase angles larger than $\alpha>40^{\circ}$, the derived fit defined in \cite{1990Zappala} is non-linear. The cumulative distribution of corrected lower limit amplitudes using both the linear (Equation~\ref{eq:phase}) and non-linear fits \citep{1990Zappala} are consistent across the observed phase angle range of $3^{\circ}<\alpha<84.5^{\circ}$. The cumulative distribution of lower limit amplitudes using the linear fit is shown in Figure~\ref{fig:DeltaV} in orange. 

\cite{2019McNeill} finds a Gaussian shape distribution for the NEO population with a mean axis ratio of $\frac{b}{a}=0.72\pm0.08$, describing the elongation with respect to spherical ($b=a$). From this distribution, we use a Monte Carlo model to generate peak to trough lightcurve amplitudes ($A$) following Equation~\ref{eq:Amp} below. We assign object obliquity values following the distribution constrained by \cite{2017Tardioli} which leverages the Yarkovsky effect and its dependence on an asteroid's obliquity.

\begin{equation} \label{eq:Amp}
A=-2.5\log\frac{b}{a}
\end{equation}

The derived cumulative fraction of amplitudes using this method is shown in Figure~\ref{fig:DeltaV} in blue. For the 48 high albedo targets (23 DCT/Calar Alto, 25 LCDB), we find a preference for less elongated objects when compared to the NEO shape distribution \cite{2019McNeill}. We also find three potential candidates for extreme elongated objects: 1865 (1971 UA), 416186 (2002 TD60), and 2010 NR1. These are objects which have a phase corrected lightcurve amplitude greater than 1~magnitude \citep{2018McNeill}. Because these are lower limit amplitudes, there may be other candidates in our sample, and so we cannot rule out highly elongated objects as a contributor to the factor of 8 increase in high albedos ($p_V\ge 0.5$) derived in the \spitzer~2017 dataset.

% Extreme elongated candidates: 2010 NR1, 2002 TD60, 1971 UA

%%%%%%%%%%%%%%%%%%%%%%%%%%%%%%%%%%%
%%%%%%%  				METHODS 			%%%%%%
%%%%%%%%%%%%%%%%%%%%%%%%%%%%%%%%%%%

\section{Modeling} \label{sec:MC1}

There are many possible contributors to the unexpectedly high albedo values including rotational lightcurve variability (leading to a misestimate of asteroid diameter), inaccurate absolute visual magnitudes (either from poor photometry or lightcurve effects), or effects from non-simultaneous observations due to changes in observing geometry, phase angle, and rotational phase. We use Monte Carlo (MC) simulations to explore albedo offsets as a result of these factors.

%%%%%%%%%%%%%%%%%%%%%%%%%%%%%%%%%%%

We perform a MC simulation on all \spitzer~targets to assess the effects of amplitude on the resulting albedo distribution of our dataset. We select a peak to trough lightcurve amplitude in the range of 0.1~magnitudes to 1.6~magnitudes. We assume that all targets in our sample have a sinusoidal lightcurve of the same amplitude. We also assume that the lightcurve variability is consistent between the visible and thermal wavelengths. 

The following steps are applied in the MC simulation:

\begin{enumerate}
\item Given a lightcurve amplitude, we randomly sample a rotational phase for each target from a uniform distribution of $H' = H \pm A/2$ where $H$ is the corrected AstDyS H-magnitude and $A$ is the peak to trough lightcurve amplitude we have assigned for all targets. By sampling the rotational phase, we simulate the uncertainties associated with observing partial lightcurves. 

\item Using $H'$ for each target found in Step~1, we apply a photometric uncertainty by randomly sampling from a normal distribution of $H' \pm \sigma$ where $\sigma$ is the photometric error provided for the corrected AstDyS H-magnitude of the target. The randomly selected value is $H''$.

\item Steps~1 and 2 are similarly applied to the thermal lightcurve where flux ($f$) and $\sigma$ are from the \spitzer~measurements. Using the equivalent lightcurve variability for the thermal as visible wavelengths, we select a rotational phase and corresponding flux ($f'$) from a uniform distribution, then apply a photometric uncertainty to $f'$ from a normal distribution to select $f''$.

\item We create a distribution of slope parameters ($G$) using the \cite{2019Binzel} taxonomy distribution for NEOs and the \cite{2015Veres} mean $G$ values for each taxonomy. For each target, we randomly sample a $G$ value from a distribution of 10,000 NEOs shown in Figure~\ref{fig:Gdist}. 

\begin{figure}
  \includegraphics[width=3.4in]{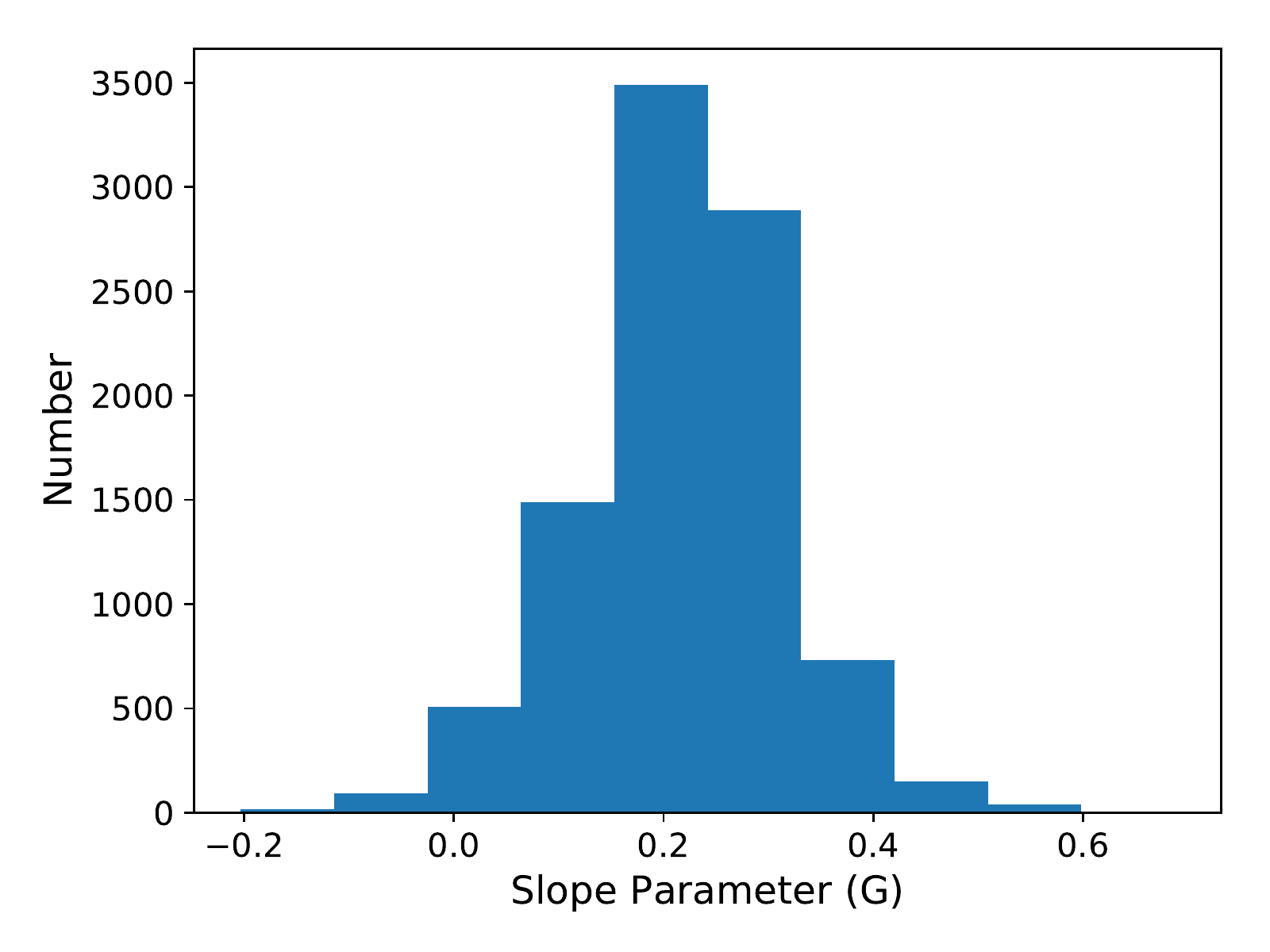}
  \caption{Distribution of G values from \cite{2015Veres} with the \cite{2019Binzel} NEO taxonomy distribution. The distribution is randomly sampled as input for the MC simulation. \label{fig:Gdist}}
\end{figure}

\item The heliocentric distance ($r$), geocentric distance ($\Delta$), and phase angle ($\alpha$) for the asteroid are obtained from the NASA JPL Horizons Database\footnote{\url{https://ssd.jpl.nasa.gov/horizons.cgi}} at the midpoint time of the \spitzer~observation. The resulting H-magnitude ($H''$), thermal flux ($f''$), slope parameter ($G$), and observing geometry of the target ($r, \Delta, \alpha$) are used as inputs in the modified NEATM \citep{2010Trilling, 2016Trilling} to generate an albedo for each target. 

\item \cite{2010Trilling} and \cite{2016Trilling} account for $\eta$ uncertainties in a MC approach in the NEATM that varies $\eta$. The resulting uncertainties due to $\eta$ are reflected in the upper and lower 1--$\sigma$ published albedo uncertainties. Our observations show no preference for high albedo objects observed at high phase angle (Figure~\ref{fig:pv_phase}). As a result, we can treat uncertainties due to $\eta$ equally for all targets. We create a distribution of the 1--$\sigma$ upper and lower albedo uncertainties for the targets provided by \cite{2010Trilling, 2016Trilling}. We randomly select an upper and lower uncertainty, and use the mean of those values as sigma to create a normal distribution about the NEATM-derived albedo from Step~5. Our final albedo value is sampled from this normal distribution of $p_V\pm\sigma$. 

\begin{figure}
  \includegraphics[width=3.4in]{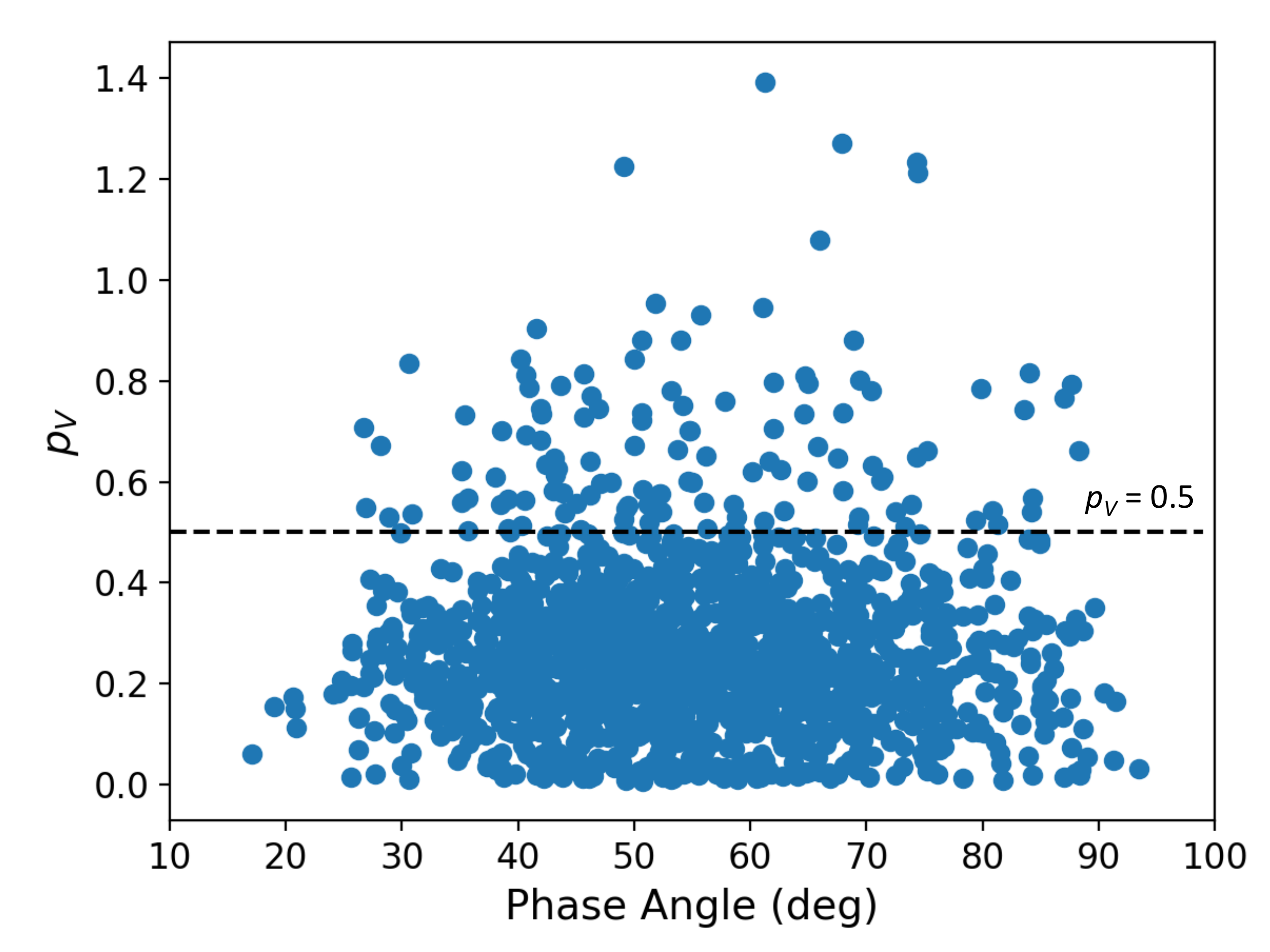}
  \caption{Albedo vs.\ Phase Angle for 1505 \spitzer~2017 targets. There is no preference for high albedo objects observed at high phase angle. \label{fig:pv_phase}}
\end{figure}

\end{enumerate}

We consider this technique for three different definitions of high albedo thresholds (0.4, 0.5, 0.6). 

%%%%%%%%%%%%%%%%%%%%%%%%%%%%%%%%%%%
%%%%%%				RESULTS			%%%%%%%%
%%%%%%%%%%%%%%%%%%%%%%%%%%%%%%%%%%%

\section{Results} \label{sec:results}

\begin{figure}
  \includegraphics[width=3.4in]{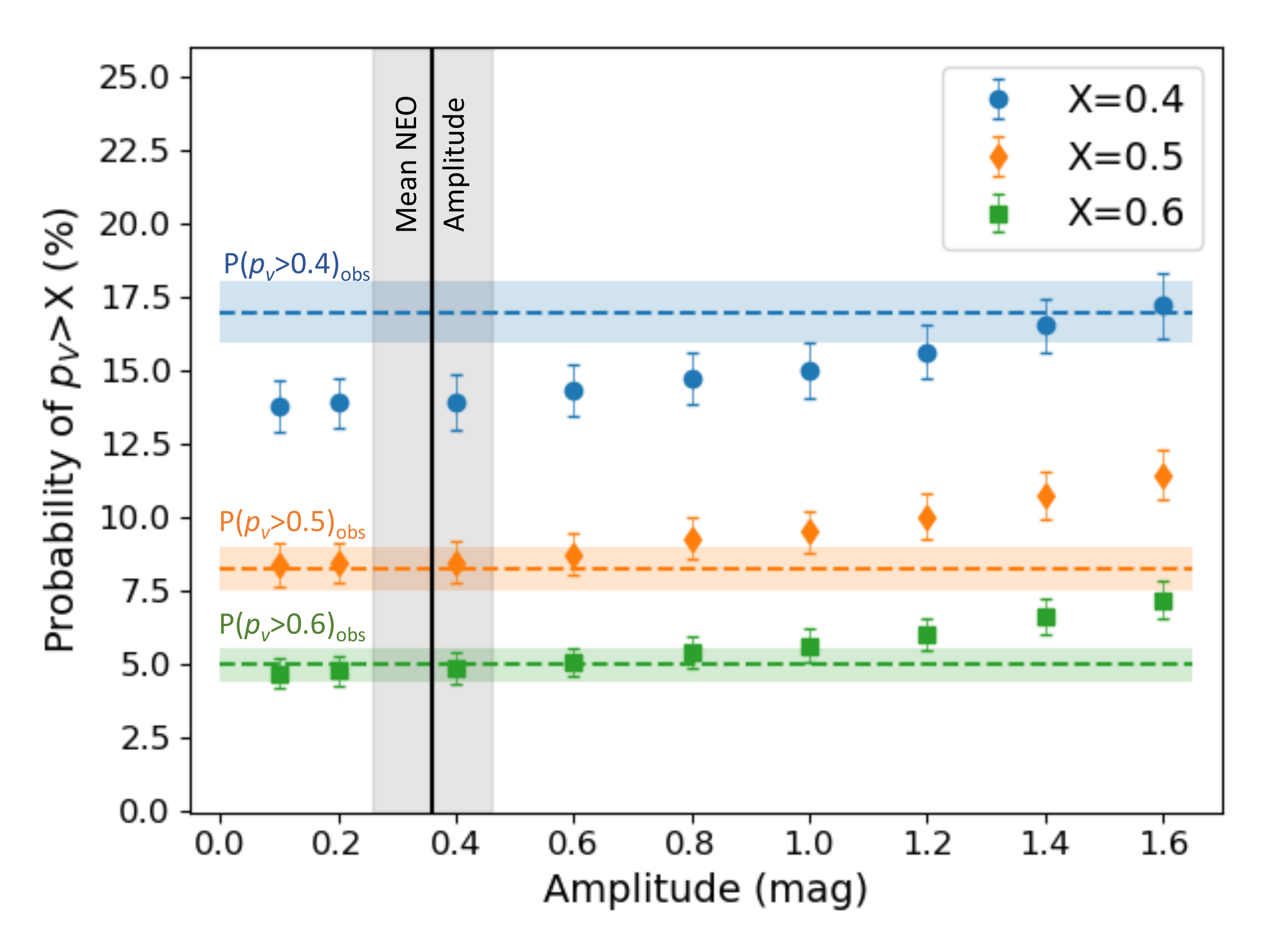}
  \caption{Probability for targets to have derived high albedo (0.4, 0.5, 0.6) as a function of lightcurve amplitude, as derived from our MC model. The three observational thresholds are indicated as horizontal dashed lines with 1--$\sigma$ shaded uncertainties. The MC probability distribution is plotted as individual data points with 1--$\sigma$ uncertainties for each high albedo threshold at different lightcurve amplitudes. The \cite{2019McNeill} mean lightcurve amplitude is shown as the vertical black line with 1--$\sigma$  uncertainties. We see that for the 0.5 and 0.6 albedo threshold, we can match the observations at the mean NEO lightcurve amplitude. However, for the 0.4 albedo threshold, there is an excess of targets in the observations with respect to the MC derived fraction.  \label{fig:MC_amp}}
\end{figure}

Figure~\ref{fig:MC_amp} shows the overall fraction of targets with high albedo (0.4, 0.5, 0.6) as a function of input amplitude, as derived from our MC model. From the observations, $\sim$17\% of \spitzer~targets have albedo values greater than 0.4,  $\sim$8\% have albedos greater than 0.5, and $\sim$5\% have albedos greater than 0.6. These three observational thresholds are indicated as horizontal dashed lines with Poisson uncertainties ($\frac{\sqrt{N}}{N-1}$) shown as the shaded regions around the dashed line. We use MC simulations to find the probability that a \spitzer~target with a given lightcurve amplitude (horizontal axis) has a MC derived albedo value greater than a given high albedo threshold (0.4, 0.5, 0.6). The MC probability distribution is plotted as individual data points with 1--$\sigma$ uncertainties for each lightcurve amplitude. The color of the data points matches the threshold value in consideration. 

We find a large fraction of \spitzer~targets at the albedo threshold 0.4. As a result, we require large lightcurve amplitudes consistent with extreme elongated objects ($>1$~magnitude) to describe the observations with shape alone. Thresholds of 0.5 and 0.6 are consistent with the observations at the mean lightcurve amplitude (vertical black line) derived for the NEO shape distribution in \cite{2019McNeill}. These results lead us to conclude the following: 

\begin{enumerate}

\item We require NEOs in our sample to have intrinsic albedos up to 0.5. For $p_V<0.5$, we find a larger Observed fraction than the MC fraction, implying a compositional influence on albedo rather than lightcurve effects. 

\item We match the Observed fraction of targets with high albedo ($p_V\ge 0.5$) in the \spitzer~2017 dataset at lightcurve amplitudes consistent with the mean amplitude of the NEO shape distribution (Figure~\ref{fig:MC_amp}). When we adopt the \cite{2019McNeill} shape distribution for lightcurve amplitude, we place an upper limit on intrinsic albedo in the NEO population of $0.5\pm0.1$. Thus, high albedo targets are not required in our dataset. 

\item We can explain the factor of 8 increase in targets with $p_V>0.5$ as a result of lightcurve sampling and thermal model uncertainties.  

\end{enumerate}

%%%%%%%%%%%%%%%%%%%%%%%%%%%%%%%%%%%
%%%%%%			DISCUSSION			%%%%%%%%
%%%%%%%%%%%%%%%%%%%%%%%%%%%%%%%%%%%

\section{Investigation into Intrinsically\\ High Albedo NEOs}  \label{sec:HA_NEOs}

While our result does not require intrinsically high albedo NEOs ($p_V>0.5$), it also does not exclude them. We investigate the need for high albedo NEOs by adopting the \cite{2019McNeill} shape distribution described in Section~\ref{sec:Phot}, and employ a MC technique. For each target, we select $A$ from the NEO shape distribution of amplitudes for both the visible and thermal lightcurve. We repeat Steps 1--6 described in the previous MC (Section~\ref{sec:MC1}).

\begin{figure}    
  \includegraphics[width=3.4in]{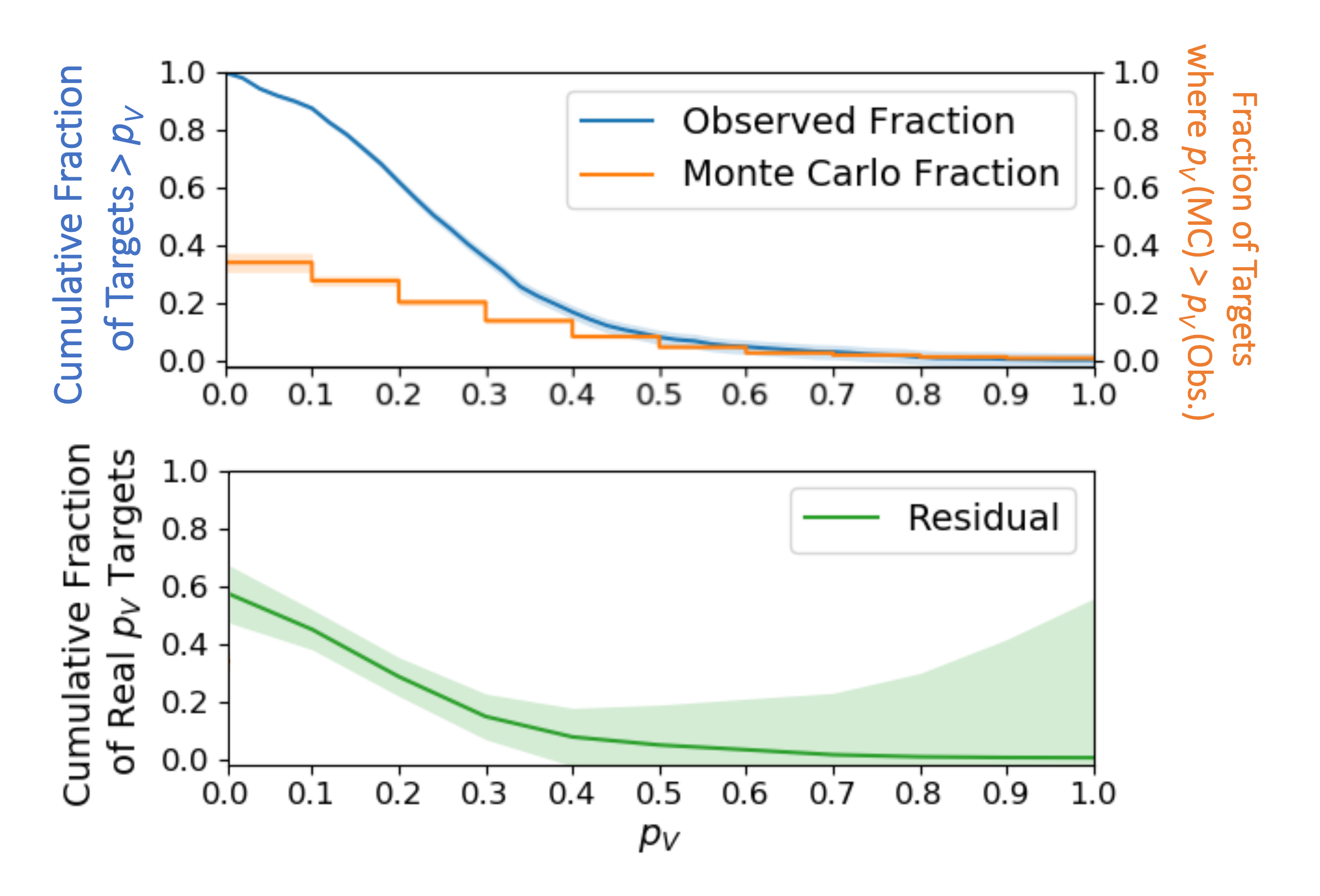}
  \caption{Top: The Observed fraction is a cumulative distribution function of the 1505 \spitzer~2017 targets for albedo values from 0.1--1.0 (horizontal axis). The MC fraction represents the fraction of targets with albedo in a 0.1 bin width (i.e., 0.1--0.2, 0.2--0.3, etc.) that move out of their published albedo bin to a larger bin due to fluctuations from the lightcurve and diameter uncertainties sampled in the MC model. The 1--$\sigma$ uncertainties are shown for both trend lines as shaded regions. Bottom: The residual is the cumulative fraction of observed targets with the MC fraction of targets removed. That is, the lightcurve and diameter fluctuations which produce a large nominal albedo are removed from the sample. The result is the fraction of targets in the \spitzer~dataset with real albedo values, finding a maximum nominal albedo of 0.5$\pm$0.1. The 1--$\sigma$ uncertainties shown as a shaded region. We find an upper limit on albedo of $0.5\pm0.1$ where the Residual is consistent with zero for albedo bins greater than 0.5. For the high albedo bins, the fluctuations are high and we find all targets leave their starting bins. This, however, does not exclude targets from having intrinsically high albedos. \label{fig:HA}} 
\end{figure}

We compare this derived MC albedo to the published \spitzer~2017 albedo for the target and find the fraction of targets with MC albedo greater than the published \spitzer~2017 albedo. The top panel of Figure~\ref{fig:HA} shows the cumulative distribution function of albedo for the \spitzer~published values in blue -- what we call the Observed fraction. Poisson uncertainties ($\frac{\sqrt N}{N-1}$) are represented as a shaded region around the data. In orange, we plot the mean fraction of targets from our MC trials with MC albedo values larger than that of their published \spitzer~albedo. We only consider the cases where the MC albedo falls in a higher 0.1--width albedo bin. An example is a \spitzer~2017 target with albedo of 0.57 (0.5--0.6 albedo bin) and a MC derived albedo for that target of 0.62 (0.6--0.7 albedo bin). The MC derived albedo is in a larger albedo bin, so we consider this target in our analysis. We expect there to be equally as many cases where the MC albedo is less than the published \spitzer~2017 albedo, however, we are only interested in situations which produce a high nominal albedo. The situations which produce a low nominal albedo fall out of the scope of this work and are therefore not considered. In Figure~\ref{fig:HA}, the 1--$\sigma$ uncertainties from all MC trials are displayed as a shaded region around the MC fraction. The first bin shows that $\sim 40\%$ of the targets in the sample that have published albedo values in the range 0--0.1 have ``upper'' MC albedo values greater than 0.1. The MC fraction thus represents the fluctuation and uncertainty in the derived albedo value due to both lightcurve and thermal model input uncertainties used in the MC simulation.    

The bottom panel of Figure~\ref{fig:HA} shows the cumulative fraction of derived intrinsic albedo targets. We find the residual of the two datasets from the top panel for each albedo bin using Equation~\ref{eq:res}. We define the residual to be the fractional difference between the Observed fraction and the MC fraction where $f_{OBS}$ is the Observed fraction (blue) and $f_{MC}$ is the MC fraction (orange):
\begin{equation} \label{eq:res}
	Residual = f_{OBS}[1-f_{MC}]
\end{equation}

The uncertainties provided on the residual are 1--$\sigma$ fractional uncertainties from the Observed fraction and MC fraction added in quadrature. The residual represents the fraction of targets in the \spitzer~2017 dataset which we believe to have the displayed intrinsic albedo values. That is, we do not require any targets to have intrinsic albedos where the cumulative fraction goes to zero. The result found in the residual is consistent with previous work by \cite{2019Binzel} and \cite{2011Thomas}. We can place an upper limit on albedo in the NEO population of 0.5$\pm$0.1 beyond which high albedo targets are not necessary to describe our observations. The precision in our upper limit is limited by our sampling size of 0.1--width albedo bins. However, the residual is formally nonzero to arbitrarily large values. 
 
\begin{figure}
  \includegraphics[width=3.4in]{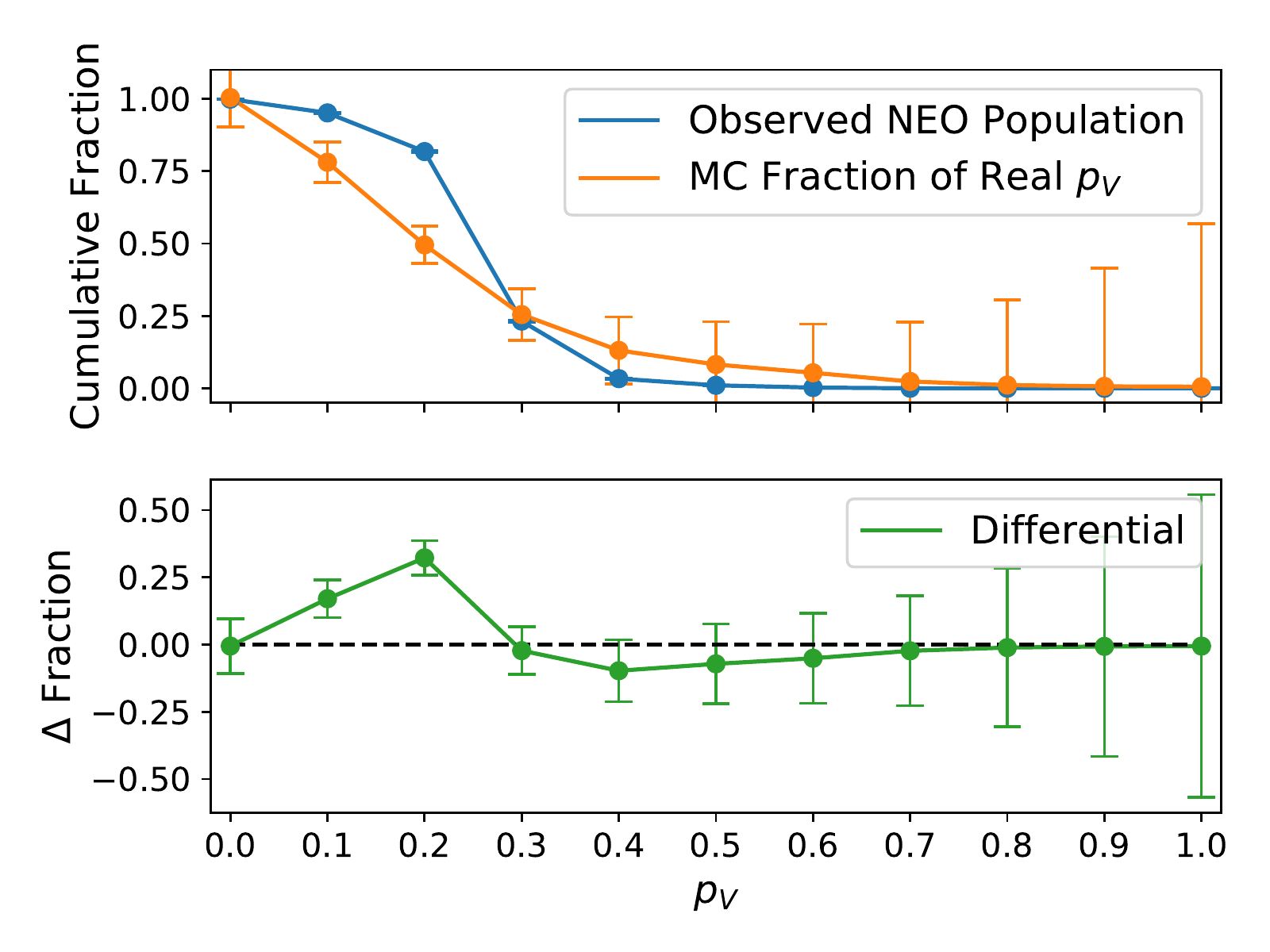}
  \caption{Top: The cumulative albedo distribution for the NEO population from \cite{2011Thomas} and \cite{2019Binzel} shown in blue. For comparison, we plot the normalized cumulative MC fraction of real albedo targets (orange). The albedo distributions are consistent for albedo greater than 0.3, but we see fewer low albedo targets in the \spitzer~dataset compared to the NEO population. Bottom: The difference between the Observed fraction (blue) and the MC fraction (orange). The \spitzer~observations are biased toward higher albedos and lack a range of low albedo asteroids. The Differential in the bottom panel may be reflective of this bias. \label{fig:NEOPop}}
\end{figure}

We explore the possibility that high albedo targets exist in the range $0.5\le p_V\le0.7$, and may as a result, be interesting due to their unique shape or composition. We compare our MC fraction of real albedo targets (Residual; Figure~\ref{fig:HA}) to the albedo distribution of the known NEO population \citep{2011Thomas,2019Binzel}. We normalize the cumulative MC fraction of real albedo targets to 100\% for $p_V>0$, treating this as a standalone population. The two distributions are shown in Figure~\ref{fig:NEOPop}. 

The observed distribution is determined from \cite{2019Binzel} ground-based observations of the NEO taxonomy distribution and mean albedo values for each taxonomy defined by \cite{2011Thomas}. We find that the cumulative fraction of MC targets is consistent within 1--$\sigma$ for $p_V\ge0.3$. However, there is a excess of low albedo MC targets ($p_V<0.3$) compared to the NEO population which may be a result of objects appearing brighter in the infrared than the visible. This bias may contribute to the trend seen in the Differential (Figure~\ref{fig:NEOPop}; bottom). 

\begin{figure}
  \includegraphics[width=3.4in]{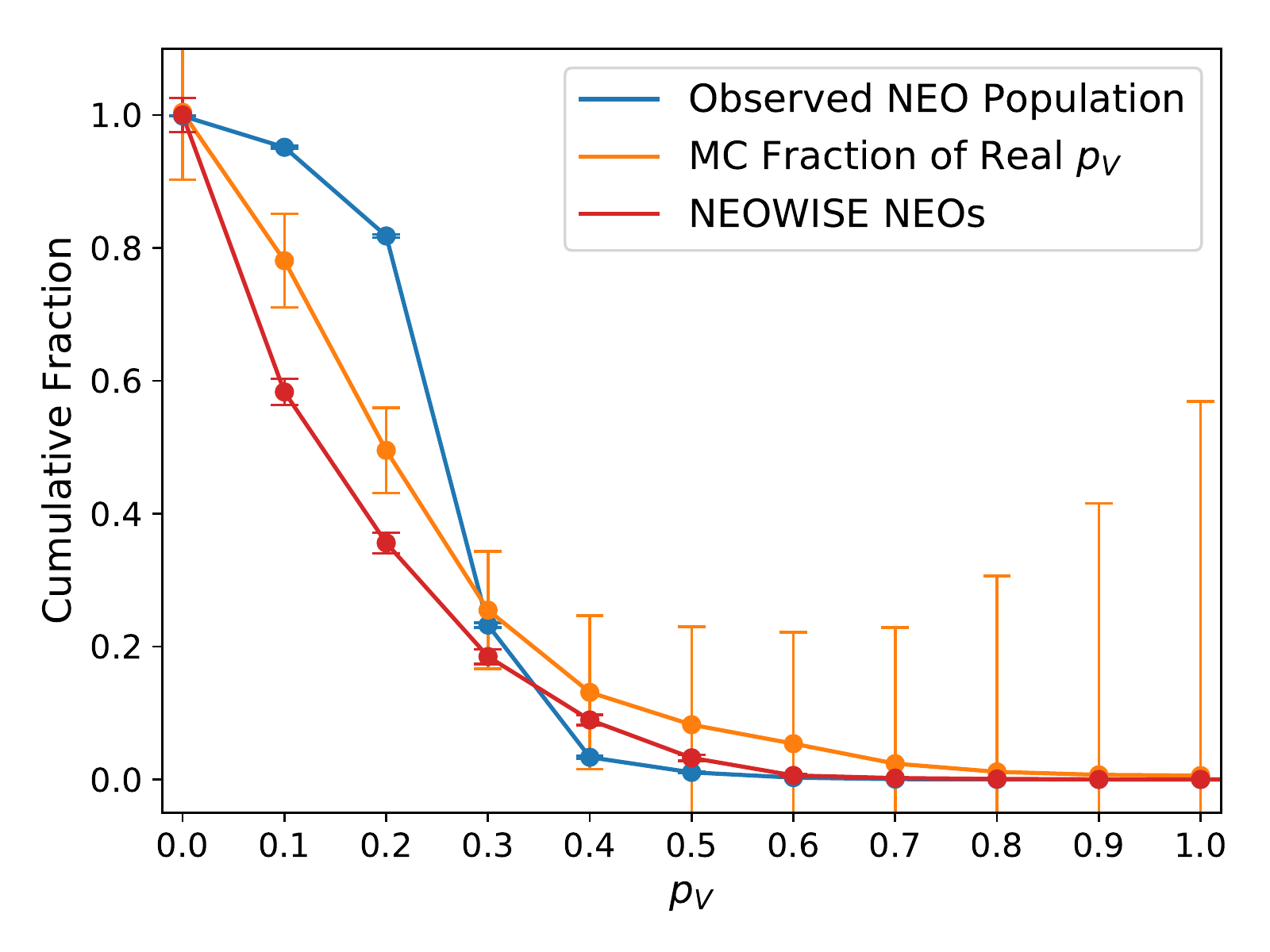}
  \caption{The cumulative albedo distribution for the observed NEO population \citep{2011Thomas, 2019Binzel} shown in blue, the MC fraction shown in orange, and the NEOWISE dataset \citep{2019Mainzer} shown in red. The albedo distributions for the MC fraction and the NEOWISE dataset are similar with relatively high fractions of both low and high albedo targets. These trends are unlike the Observed NEO Population which has most of its targets in the albedo range of 0.2-0.4.\label{fig:NEOWISE}}
\end{figure}

In Figure~\ref{fig:NEOWISE}, we include the cumulative fraction of NEOWISE albedos for NEOs published in \cite{2019Mainzer}. Because NEOWISE operates in survey mode, they are only biased toward detectability and not in target selection. As a result, we expect a higher fraction of low albedo objects in the NEOWISE sample than is present for the optically-selected \spitzer~sample and the observed NEO population. This is consistent with the trend we observe in Figure~\ref{fig:NEOWISE} where $\sim80\%$ of the NEOWISE sample has an albedo less than 0.3.

We have shown Figure~\ref{fig:HA} that we have targets in our \spitzer~2017 dataset with true albedos up to 0.5. For the purpose of this investigation, we will explore those with potentially rare high albedos ($p_V\ge 0.5$) where both shape and composition play a role in influencing their unexpected high albedo values. 

In the \spitzer~2018 sample from ExploreNEOs, NEOSurvey, and NEOLegacy of 1999 targets as of JD 2458278 (2018-06-08), 181 (9\%) possibly have intrinsic high albedo values. We compare the taxonomy distribution of the 181 high albedo \spitzer~2018 targets with that of the known NEO population \citep{2019Binzel, 2011Thomas} in an effort to disentangle shape influences from compositional influences on albedo. We adopt the taxonomic complexes defined in \cite{2011Thomas} as shown in Table~\ref{tab:tax_complex}. We hypothesize that a taxonomic distribution of these high albedo targets will favor complexes with higher intrinsic mean albedos (X-, R-, and V-types) if we have intrinsic high albedo outliers in our sample. Of these three taxonomy complexes, observations of the NEO population find R-types are the least common ($<1\%$), V-types represent $\sim4\%$, and X-types $\sim8\%$ \citep{2019Binzel}. Of the X-type complex, Xe- and E-types are known to have the highest mean albedo at $p_V = 0.536\pm0.247$ \citep{2013DeMeo}. 

\begin{deluxetable}{ccccc}
\tablecaption{Taxonomy Complex \label{tab:tax_complex}}
\tablecolumns{5}
\tablenum{5}
\tablewidth{0pt}
\tablehead{
\colhead{Complex}	& \colhead{Types}	& \colhead{$p_V$}	& \colhead{$\sigma^+$}	& \colhead{$\sigma^-$}
}	
\startdata
D	& D, T	& 0.02	& 0.02	& 0.01 \\
C	& B, C, Cb, Cg, Ch, Cgh	& 0.23	& 0.15	& 0.13 \\
S	& S, Sa, Sk, Sl, Sr, K, L, Ld	&0.26 	& 0.04	& 0.03 \\
Q	& Q, Sq	& 0.29	& 0.05	& 0.04 \\
X	& X, Xc, Xk, Xe, E, M, P & 0.31	&0.08 	&0.07 \\
R	& R	&*0.340 	&*0.034 	&*0.034 \\
V	& V	& 0.42	&0.13 	& 0.11 
\enddata
\tablecomments{Types included in each taxonomy complex with average albedo values from \cite{2011Thomas} and *\cite{2004Stuart}.}
\end{deluxetable}

\begin{figure}
  \includegraphics[width=3.4in]{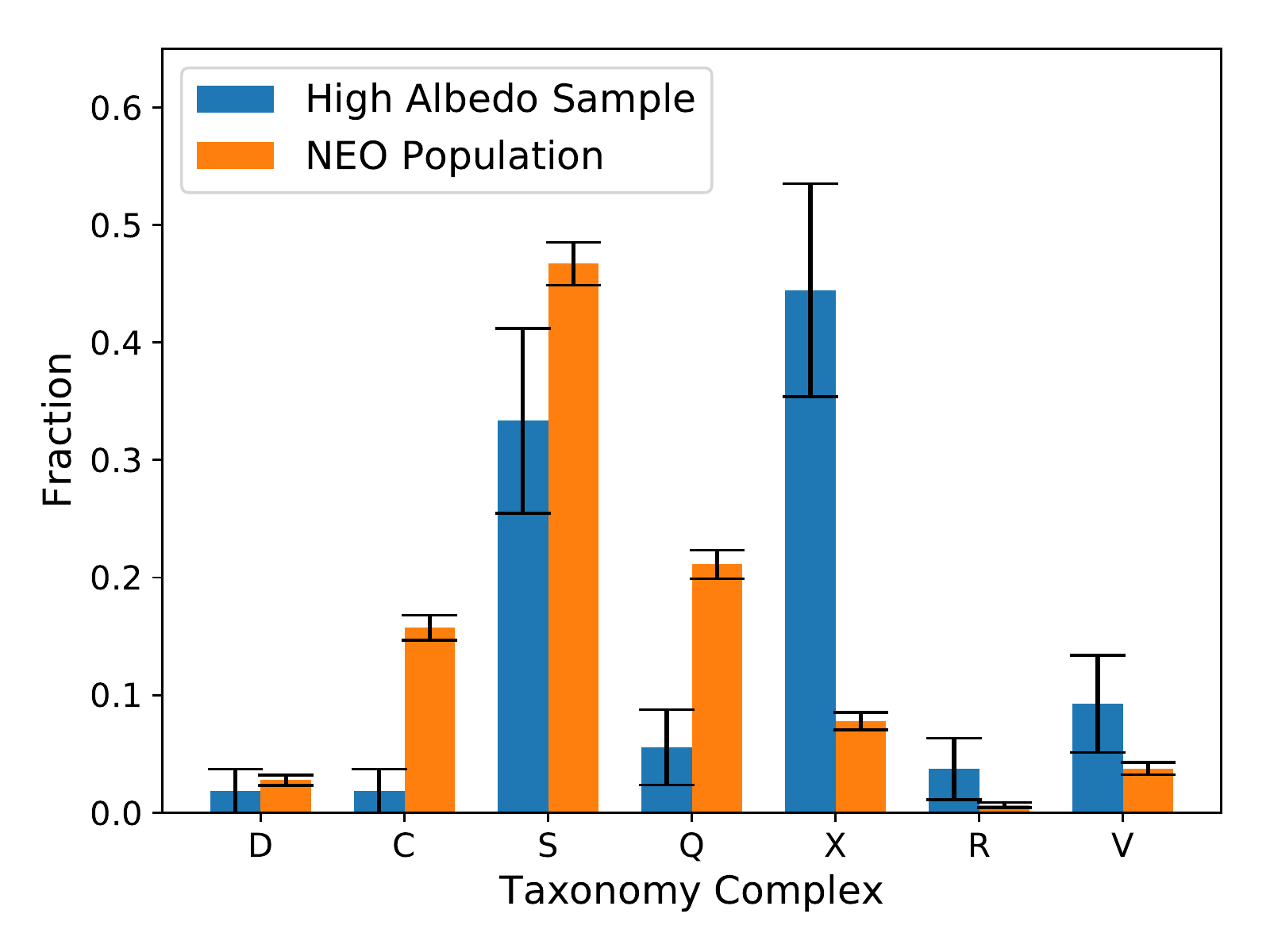}
  \caption{The NEO taxonomy distribution is shown in order of increasing mean taxonomic albedo \citep{2011Thomas} from left to right using NEO taxonomy data from \cite{2019Binzel}. The 48 high albedo targets with classified taxonomies are distributed over all represented taxonomic complexes in the NEO population. We find evidence for higher nominal mean albedo taxonomies (X-, R-, V-types) and less of a preference for D-, C-, and S-, and Q-types. The strongest excess is in the X-types which we believe is due to the presence of high albedo Xe-types in our sample. \label{fig:HA_TAX}}
\end{figure}

Of the 181 \spitzer~2018 targets with nominal albedo values greater than 0.5, 54 (30\%) have measured taxonomies \citep{2002Binzel, 2011Thomas, 2014Thomas, 2016Carry, 2017Erasmus, 2018Perna}. Table~\ref{tab:HA_taxonomy} shows a complete list of these targets. A comparison of the taxonomy distributions of these high albedo targets with the NEO population is shown in Figure~\ref{fig:HA_TAX} in order of increasing mean albedo. We find an excess of X-, R-, and V-type complexes in the high albedo sample with respect to the NEO population, with only the X-type fraction significantly greater than the NEO population (Figure~\ref{fig:HA_TAX}). 30\% of the classified high albedo X-types have been identified as E-types \citep{2011Thomas}, supporting a compositional-driven influence for a fraction of the targets in this high albedo sample. While the R- and V-type fractions are within 1--$\sigma$ of the NEO population, there is a large excess of X-types in the high albedo sample, most likely a result of Xe- and E-types. While the Xe- \citep{2009DeMeo} and E-types \citep{1984Tholen} are defined using different classifications, they are nearly identical \citep{2019Binzel}. From here on, we will refer to Xe- and E-types together as Xe. 

\cite{2019Binzel}, \cite{2018Granvik}, and \cite{1992Gaffey} all find a strong signature for the inner Main Belt Hungaria region as a source for the Xe-type NEOs. Xe-types show strong mineralogic similarity to the enstatite-rich aubrite meteorites \citep{2018Lucas} with mean albedos greater than 0.5 \citep{2013DeMeo}. In this high albedo sample, we also find representation from D-, C-, S-, and Q-type taxonomic complexes. There is only 1 object each of D-type \citep{2016Carry} and C-type \citep{2001Whiteley} taxonomies of the lowest mean albedos classified with spectrophotometry. We believe these objects may belong to the X-type complex based on their extreme nominal high albedos, but this has yet to be confirmed spectroscopically. D-, C-, and X-types have similar featureless red sloped spectra in the visible to near-infrared, but with varying degrees of spectral reddening, so it is not unreasonable for them to be misclassified due to the similarity in color between the three types. It is, however, unreasonable for an object with high albedo to be classified as D- or C-type. 

We find a similar fraction of S-type objects in our high albedo sample in comparison to the NEO population where we would have expected a deficit. For an S-type object with albedo 3--$\sigma$ above the mean, we require a lightcurve amplitude of at least 0.5 and a poor $H$ measurement to derive an albedo greater than 0.5. As a result, it is possible that a fraction of the S-type objects in our sample produce nominal high albedo values due to influence from both shape and composition.

%%%%%%%%%%%%%%%%%%%%%%%%%%%%%%%%%%%
%%%					SUMMARY	 		 %%%%%
%%%%%%%%%%%%%%%%%%%%%%%%%%%%%%%%%%%
\section{Summary}

Major infrared surveys find that the fraction of NEOs with high albedos is a factor of 8 more than expected based on average albedos from the literature and the known NEO taxonomy distribution. This excess in high albedo values is easily explained by offsets in photometry introduced due to incomplete lightcurve sampling and uncertainties on remaining thermal model input parameters including the beaming parameter. With full visible lightcurves, the H-magnitudes can be improved and uncertainties reduced, but without a full thermal lightcurve, we cannot derive accurate intrinsic albedos. For a sample of 48 high albedo NEOs with lower limit amplitudes, we do not find that their shapes are significantly different from the \cite{2019McNeill} NEO shape distribution. We adopt the McNeill NEO amplitude distribution in a MC model sampling uncertainties on thermal model input parameters and find that the upper limit on albedo in the NEO population is $0.5\pm0.1$.   

There is a significant fraction of Xe-type objects, which have high albedos, present in our catalog of high albedo objects, and perhaps, an excess of X-, R-, and V-type complexes above the NEO population taxonomy distribution. However, the mean albedo for each of these taxonomies all lie near or under the 0.5 high albedo threshold. As a result, these findings are consistent with previously published works which also find an upper limit albedo in the NEO population of 0.5.   

\startlongtable
\begin{deluxetable*}{cccccc}
\tablecaption{High Albedo Taxonomy \label{tab:HA_taxonomy}}
\tablecolumns{6}
\tablenum{5}
\tablewidth{0pt}
\tablehead{
\colhead{\#}	& \colhead{Designation} 	&\colhead{$p_V$}	&\colhead{$\sigma^+$} 	&\colhead{$\sigma^-$}	& \colhead{Taxonomy Reference}
}		
\startdata
\multicolumn{6}{c}{\it{C-Complex:}}			\\
152637 		&1997 NC1		& 0.541			& 0.228				& 0.231				& [25] \\ 
\hline
\multicolumn{6}{c}{\it{D-Complex:}}			\\
225312 		&1996 XB27		& 0.544 			&0.289 				&0.261				& [7] \\ 
\hline
\multicolumn{6}{c}{\it{Q-Complex:}}			\\
23183 		&2000 OY21		& 0.904 (0.640)		&0.404 (0.076)			&0.426 (0.076)		 	& [8]  \\ 
136849 		&1998 CS1		&0.504			&0.356				&0.298				& [22] \\ 
341816 		&2007 YK			& 0.75			& 0.432				& 0.393				& [19]	\\
\hline
\multicolumn{6}{c}{\it{R-Complex:}}			\\
257838 		&2000 JQ66		&0.673			& 0.452				& 0.399				& [3] \\ 
482566		&2012 WK4		& 0.612			&0.276 				&0.269				&[20] \\
\hline
\multicolumn{6}{c}{\it{S-Complex:}}			\\
1620 		&1951 RA			& 0.560 (0.290)		& 0.354 (0.038) 		& 0.316 (0.038)		 	& [6] \\ 
1865 		&1971 UA			& 0.578 (0.4797)	& 0.311 (0.4074)		& 0.289 (0.4074)		 & [3,6,24] \\ 
1915 		&1953 EA			& 0.535			&0.303				& 0.274				&[1,6] \\ 
3199 		&1982 RA			& 0.531			&0.379				&0.324				& [6] \\ 
5131 		&1990 BG			& 0.834			&0.497				&0.470				& [4] \\ 
10165 		&1995 BL2		& 0.567 (0.372)		&0.353 (0.198)			&0.309 (0.198) 			& [3,6] \\ 
25143 		&1998 SF36		& 0.496			&0.222				&0.210 				& [2,7] \\ 
42286 		&2001 TN41		& 0.548			&0.422				&0.338				& [11] \\ 
87024 		&2000 JS66		& 0.669			&0.280				&0.289				& [5] \\
153953 		&2002 AD9		& 0.556			&0.350				&0.309				& [3,14] \\ 
159609		&2002 AQ3		& 0.736			&0.336				&0.342				& [9,21] \\ 
200754 		&2001 WA25		& 0.501 (0.21)		&0.208 (0.204)			&0.209 (0.204)  		& [3,8,18] \\ 
200840 		&2001 XN254		& 0.646			&0.262				&0.281				& [5]  \\
252399 		&2001 TX44		& 0.647 (0.478)		&0.338 (0.157)			&0.316 (0.157) 			 & [3] \\ 
274855 		&2009 RB4		& 0.539			&0.304				&0.277				& [5] \\
416186 		&2002 TD60		& 0.496			& 0.294				&0.262				& [17] \\ 
			&2005 ML13		& 0.516			&0.235				&0.216				& [20] \\ 
			&2015 JD1		& 0.507			&0.307				&0.247				& [20] \\ 
\hline
\multicolumn{6}{c}{\it{V-Complex:}}			\\
54509 		&2000 PH5		& 0.556			&0.272				&0.254				& [3,12] \\ 
137052 		&1998 VO33		& 0.621			&0.374				&0.338				& [3,9] \\ 
470975 		&2009 SC15		& 0.620			&0.281				&0.275				& [15] \\ 
			&1999 SH10		& 0.523			& 0.239				&0.223				&  [10] \\
			&2014 HR178		& 0.801			&0.347				&0.330				& [5] \\
\hline
\multicolumn{6}{c}{\it{X-Complex:}}			\\
3671 		&1984 KD			& 0.625			&0.262				&0.277				& [4,6] \\ 
*3691 		&1982 FT	 		& 0.537 (0.593)		&0.368 (0.120)			&0.311 (0.120)		 	& [4,6] \\ 
*5751 		&1992 AC	 		& 0.497			&0.306				&0.265				& [4,16] \\
7474 		&1992 TC			& 1.225			&0.490				&0.543				& [3,9] \\ 
*10302 		&1989 ML 		& 0.559			&0.274				&0.251				& [2,3] \\ 
*17511 		&1992 QN 		& 0.766			&0.249				&0.279				& [4,14] \\ 
*33342 		&1998 WT24 		& 0.816 (0.654)		&0.261 (0.13)			&0.292 (0.13)			& [18] \\ 
*54789 		&2001 MZ7 		& 0.556 (0.856)		&0.277 (0.145)			&0.275 (0.145)			& [8,24] \\ 
*136564 		&1977 VA 			& 0.584			&0.287				&0.280				& [3] \\
136818		&1997 MW1		& 0.781			&0.277				&0.315				& [2,25] \\ 
138893 		&2000 YH66		& 0.693			&0.343				&0.343				& [3] \\ 
162004		&1991 VE			& 0.699			& 0.271				& 0.298 				& [19] \\
162581 		&2000 SA10		& 0.564			&0.250				&0.245				& [14] \\
163132		&2002 CU11		& 0.809 (0.408)		&0.301 (0.061)			&0.321 (0.061) 			 & [24] \\ 
170086 		&2002 XR14		& 0.512			&0.221				&0.219				& [9] \\ 
275792 		&2001 QH142		& 0.814			&0.347				&0.368				& [24] \\ 
376848 		&2001 RY47		& 0.818			&0.295				&0.315				& [17] \\ 
407656 		&2011 SL102		& 0.577			&0.281				&0.276				& [23] \\
412976 		&1987 WC		& 0.498			&0.278				&0.242				& [5] \\
416151 		&2002 RQ25		& 0.792			&0.277				&0.286				& [24] \\ 
462559		&2009 DD1		& 0.783			& 0.383				& 0.355				& [20]\\
			&2005 TS15		& 0.728			&0.405				&0.357				& [5] \\
			&2012 PG6		& 0.496			& 0.297				& 0.247				& [20]\\
			&2014 YD			& 1.102			& 0.428				& 0.418				& [10]
\enddata
\tablecomments{High albedo NEOs with measured taxonomy. Those objects noted with * are previously identified E-types in the X-type complex. The \spitzer~albedos are listed in the second column with 1--sigma uncertainties. NEOWISE albedos from the 2019 NASA PDS release are listed in parentheses where available \citep{2019Mainzer}.}
\tablereferences{	[1]~\cite{1983Binzel},
				[2]~\cite{2001Binzel},
				[3]~\cite{2004Binzel},
				[4]~\cite{2004cBinzel},
				[5]~\cite{2019Binzel},
				[6]~\cite{2002Bus},
				[7]~\cite{2016Carry},
				[8]~\cite{2010DeLeon},
				[9]~\cite{2013DeMeo},
				[10]~\cite{2019Devogele},
				[11]~\cite{2017Erasmus},
				[12]~\cite{2007Gietzen},
				[13]~\cite{2007Harris},
				[14]~\cite{2012Hasselmann},
				[15]~\cite{2014Kuroda},
				[16]~\cite{2004Lazzarin},
				[17]~\cite{2005Lazzarin},
				[18]~\cite{2010Lazzarin},
				[19]~\cite{2018Lin},
				[20]~\cite{2018Perna},
				[21]~\cite{2013Sanchez},
				[22]~\cite{2010Somers},
				[23]~SMASS Website,
				[24]~\cite{2014Thomas},
				[25]~\cite{2001Whiteley}
				}
\end{deluxetable*}

%%%%%%%%%%%%%%%%%%%%%%%%%%%%%%%%%%%
%%%			ACKNOWLEDGEMENTS			%%%%%
%%%%%%%%%%%%%%%%%%%%%%%%%%%%%%%%%%%

\acknowledgments

This work is based in part on the observations made with the \spitzer~Space Telescope, which is operated by the Jet Propulsion Laboratory, California Institute of Technology under a contract with NASA. Support for this work was provided by NASA through an award issued by JPL/Caltech.

Some of the data utilized in this work were obtained at the Discovery Channel Telescope, operated by Lowell Observatory in Happy Jack, AZ, and at the Calar Alto Astronomical Observatory in Almer\'ia, Spain.

We acknowledge the NASA JPL Solar System Dynamics group for their online Horizons ephemeris service as well as the Asteroids Dynamic Site (AstDyS) Database. 

This work was made possible in part by the funding support of the NASA NEOO grant NNX17AH06G.

\vspace{5mm}
\facilities{Spitzer (IRAC), Discovery Channel Telescope (LMI), Calar Alto Astronomical Observatory (LDR-MKIII)}
\software{Source Extractor \citep{1996Bertin}, SCAMP \citep{2006Bertin}, Photometry Pipeline \citep{2017Mommert}}

%%%%%%%%%%%%%%%%%%%%%%%%%%%%%%%%%%%
%%%%%%% 			REFERENCES			%%%%%%%%%
%%%%%%%%%%%%%%%%%%%%%%%%%%%%%%%%%%%

\bibliography{SpitzerAlbedo_Final}

\begin{thebibliography}{}
\expandafter\ifx\csname natexlab\endcsname\relax\def\natexlab#1{#1}\fi
\providecommand{\url}[1]{\href{#1}{#1}}
\providecommand{\dodoi}[1]{doi:~\href{http://doi.org/#1}{\nolinkurl{#1}}}
\providecommand{\doeprint}[1]{\href{http://ascl.net/#1}{\nolinkurl{http://ascl.net/#1}}}
\providecommand{\doarXiv}[1]{\href{https://arxiv.org/abs/#1}{\nolinkurl{https://arxiv.org/abs/#1}}}

\bibitem[{{Bertin}(2006)}]{2006Bertin}
{Bertin}, E. 2006, in Astronomical Society of the Pacific Conference Series,
  Vol. 351, Astronomical Data Analysis Software and Systems XV, ed.
  C.~{Gabriel}, C.~{Arviset}, D.~{Ponz}, \& S.~{Enrique}, 112

\bibitem[{{Bertin} \& {Arnouts}(1996)}]{1996Bertin}
{Bertin}, E., \& {Arnouts}, S. 1996, \aaps, 117, 393,
  \dodoi{10.1051/aas:1996164}

\bibitem[{{Binzel} {et~al.}(2002){Binzel}, {Lupishko}, {di Martino},
  {Whiteley}, \& {Hahn}}]{2002Binzel}
{Binzel}, R.~P., {Lupishko}, D., {di Martino}, M., {Whiteley}, R.~J., \&
  {Hahn}, G.~J. 2002, {Physical Properties of Near-Earth Objects}, 255--271

\bibitem[{{Binzel} {et~al.}(2001){Binzel}, {Rivkin}, {Bus}, {Sunshine}, \&
  {Burbine}}]{2001Binzel}
{Binzel}, R.~P., {Rivkin}, A.~S., {Bus}, S.~J., {Sunshine}, J.~M., \&
  {Burbine}, T.~H. 2001, Meteoritics and Planetary Science Supplement, 36, A20

\bibitem[{{Binzel} {et~al.}(2004{\natexlab{a}}){Binzel}, {Rivkin}, {Stuart},
  {Harris}, {Bus}, \& {Burbine}}]{2004Binzel}
{Binzel}, R.~P., {Rivkin}, A.~S., {Stuart}, J.~S., {et~al.} 2004{\natexlab{a}},
  \icarus, 170, 259, \dodoi{10.1016/j.icarus.2004.04.004}

\bibitem[{{Binzel} {et~al.}(2004{\natexlab{b}}){Binzel}, {Rivkin}, {Stuart},
  {Harris}, {Bus}, \& {Burbine}}]{2004cBinzel}
---. 2004{\natexlab{b}}, \icarus, 170, 259,
  \dodoi{10.1016/j.icarus.2004.04.004}

\bibitem[{{Binzel} \& {Tholen}(1983)}]{1983Binzel}
{Binzel}, R.~P., \& {Tholen}, D.~J. 1983, \icarus, 55, 495,
  \dodoi{10.1016/0019-1035(83)90118-5}

\bibitem[{{Binzel} {et~al.}(2019){Binzel}, {DeMeo}, {Turtelboom}, {Bus},
  {Tokunaga}, {Burbine}, {Lantz}, {Polishook}, {Carry}, \&
  {Morbidelli}}]{2019Binzel}
{Binzel}, R.~P., {DeMeo}, F.~E., {Turtelboom}, E.~V., {et~al.} 2019, \icarus,
  324, 41, \dodoi{10.1016/j.icarus.2018.12.035}

\bibitem[{{Bowell} {et~al.}(1989){Bowell}, {Hapke}, {Domingue}, {Lumme},
  {Peltoniemi}, \& {Harris}}]{1989Bowell}
{Bowell}, E., {Hapke}, B., {Domingue}, D., {et~al.} 1989, in Asteroids II, ed.
  R.~P. {Binzel}, T.~{Gehrels}, \& M.~S. {Matthews}, 524--556

\bibitem[{{Bus} \& {Binzel}(2002)}]{2002Bus}
{Bus}, S.~J., \& {Binzel}, R.~P. 2002, \icarus, 158, 106,
  \dodoi{10.1006/icar.2002.6857}

\bibitem[{{Carry} {et~al.}(2016){Carry}, {Solano}, {Eggl}, \&
  {DeMeo}}]{2016Carry}
{Carry}, B., {Solano}, E., {Eggl}, S., \& {DeMeo}, F.~E. 2016, \icarus, 268,
  340, \dodoi{10.1016/j.icarus.2015.12.047}

\bibitem[{{Chambers} \& {et al.}(2017)}]{2017Chambers}
{Chambers}, K.~C., \& {et al.} 2017, VizieR Online Data Catalog, II/349

\bibitem[{{de Le{\'o}n} {et~al.}(2010){de Le{\'o}n}, {Licandro},
  {Serra-Ricart}, {Pinilla-Alonso}, \& {Campins}}]{2010DeLeon}
{de Le{\'o}n}, J., {Licandro}, J., {Serra-Ricart}, M., {Pinilla-Alonso}, N., \&
  {Campins}, H. 2010, \aap, 517, A23, \dodoi{10.1051/0004-6361/200913852}

\bibitem[{{Delb{\'o}} {et~al.}(2003){Delb{\'o}}, {Harris}, {Binzel}, {Pravec},
  \& {Davies}}]{2003Delbo}
{Delb{\'o}}, M., {Harris}, A.~W., {Binzel}, R.~P., {Pravec}, P., \& {Davies},
  J.~K. 2003, \icarus, 166, 116, \dodoi{10.1016/j.icarus.2003.07.002}

\bibitem[{{Delbo} {et~al.}(2015){Delbo}, {Mueller}, {Emery}, {Rozitis}, \&
  {Capria}}]{2015Delbo}
{Delbo}, M., {Mueller}, M., {Emery}, J.~P., {Rozitis}, B., \& {Capria}, M.~T.
  2015, {Asteroid Thermophysical Modeling}, 107--128

\bibitem[{{DeMeo} {et~al.}(2009){DeMeo}, {Binzel}, {Slivan}, \&
  {Bus}}]{2009DeMeo}
{DeMeo}, F.~E., {Binzel}, R.~P., {Slivan}, S.~M., \& {Bus}, S.~J. 2009,
  \icarus, 202, 160, \dodoi{10.1016/j.icarus.2009.02.005}

\bibitem[{{DeMeo} \& {Carry}(2013)}]{2013DeMeo}
{DeMeo}, F.~E., \& {Carry}, B. 2013, \icarus, 226, 723,
  \dodoi{10.1016/j.icarus.2013.06.027}

\bibitem[{{Devog{\'e}le} {et~al.}(2019, Submitted){Devog{\'e}le}, {Moskovitz},
  {Thirouin}, {Gustafsson}, {Magnuson}, {Thomas}, {Willman}, {Christensen},
  {Person}, {Binzel}, {Polishook}, {DeMeo}, {Hinke}, {Trilling}, {Mommert},
  {Burt}, \& {Skiff}}]{2019Devogele}
{Devog{\'e}le}, M., {Moskovitz}, N., {Thirouin}, A., {et~al.} 2019, Submitted,
  \aj

\bibitem[{{Erasmus} {et~al.}(2017){Erasmus}, {Mommert}, {Trilling},
  {Sickafoose}, {van Gend}, \& {Hora}}]{2017Erasmus}
{Erasmus}, N., {Mommert}, M., {Trilling}, D.~E., {et~al.} 2017, \aj, 154, 162,
  \dodoi{10.3847/1538-3881/aa88be}

\bibitem[{{Farmer}(2010)}]{2010Farmer}
{Farmer}, Jr., S.~E. 2010, Minor Planet Bulletin, 37, 165

\bibitem[{{Fazio} {et~al.}(2004){Fazio}, {Hora}, {Allen}, {Ashby}, {Barmby},
  {Deutsch}, {Huang}, {Kleiner}, {Marengo}, \& {Megeath}}]{2004Fazio}
{Fazio}, G.~G., {Hora}, J.~L., {Allen}, L.~E., {et~al.} 2004, \apjs, 154, 10,
  \dodoi{10.1086/422843}

\bibitem[{{Gaffey} {et~al.}(1992){Gaffey}, {Reed}, \& {Kelley}}]{1992Gaffey}
{Gaffey}, M.~J., {Reed}, K.~L., \& {Kelley}, M.~S. 1992, \icarus, 100, 95,
  \dodoi{10.1016/0019-1035(92)90021-X}

\bibitem[{{Gietzen} \& {Lacy}(2007)}]{2007Gietzen}
{Gietzen}, K.~M., \& {Lacy}, C.~H.~S. 2007, in Lunar and Planetary Science
  Conference, 1104

\bibitem[{{Granvik} {et~al.}(2018){Granvik}, {Morbidelli}, {Jedicke}, {Bolin},
  {Bottke}, {Beshore}, {Vokrouhlick{\'y}}, {Nesvorn{\'y}}, \&
  {Michel}}]{2018Granvik}
{Granvik}, M., {Morbidelli}, A., {Jedicke}, R., {et~al.} 2018, \icarus, 312,
  181, \dodoi{10.1016/j.icarus.2018.04.018}

\bibitem[{{Harris}(1998)}]{1998Harris}
{Harris}, A.~W. 1998, \icarus, 131, 291, \dodoi{10.1006/icar.1997.5865}

\bibitem[{{Harris} {et~al.}(2007){Harris}, {Mueller}, {Delb{\'o}}, \&
  {Bus}}]{2007Harris}
{Harris}, A.~W., {Mueller}, M., {Delb{\'o}}, M., \& {Bus}, S.~J. 2007, \icarus,
  188, 414, \dodoi{10.1016/j.icarus.2006.12.003}

\bibitem[{{Harris} \& {Young}(1989)}]{1989Harris}
{Harris}, A.~W., \& {Young}, J.~W. 1989, \icarus, 81, 314,
  \dodoi{10.1016/0019-1035(89)90056-0}

\bibitem[{{Hasselmann} {et~al.}(2012){Hasselmann}, {Carvano}, \&
  {Lazzaro}}]{2012Hasselmann}
{Hasselmann}, P.~H., {Carvano}, J.~M., \& {Lazzaro}, D. 2012, {SDSS-based
  Asteroid Taxonomy V1.1. EAR-A-I0035-5-SDSSTAX-V1.1}, NASA Planetary Data
  System

\bibitem[{{Hergenrother} \& {Whiteley}(2011)}]{2011Hergenrother}
{Hergenrother}, C.~W., \& {Whiteley}, R.~J. 2011, \icarus, 214, 194,
  \dodoi{10.1016/j.icarus.2011.03.023}

\bibitem[{{Hora} {et~al.}(2018){Hora}, {Siraj}, {Mommert}, {McNeill},
  {Trilling}, {Gustafsson}, {Smith}, {Fazio}, {Chesley}, \& {Emery}}]{2018Hora}
{Hora}, J.~L., {Siraj}, A., {Mommert}, M., {et~al.} 2018, \apjs, 238, 22,
  \dodoi{10.3847/1538-4365/aadcf5}

\bibitem[{{Knezevic} \& {Milani}(2012)}]{2012Knezevic}
{Knezevic}, Z., \& {Milani}, A. 2012, in IAU Joint Discussion, P18

\bibitem[{{Kuroda} {et~al.}(2014){Kuroda}, {Ishiguro}, {Takato}, {Hasegawa},
  {Abe}, {Tsuda}, {Sugita}, {Usui}, {Hattori}, \& {Iwata}}]{2014Kuroda}
{Kuroda}, D., {Ishiguro}, M., {Takato}, N., {et~al.} 2014, \pasj, 66, 51,
  \dodoi{10.1093/pasj/psu041}

\bibitem[{{Lazzarin} {et~al.}(2010){Lazzarin}, {Magrin}, {Marchi}, {Dotto},
  {Perna}, {Barbieri}, {Barucci}, \& {Fulchignoni}}]{2010Lazzarin}
{Lazzarin}, M., {Magrin}, S., {Marchi}, S., {et~al.} 2010, \mnras, 408, 1433,
  \dodoi{10.1111/j.1365-2966.2010.17268.x}

\bibitem[{{Lazzarin} {et~al.}(2004){Lazzarin}, {Marchi}, {Barucci}, {Di
  Martino}, \& {Barbieri}}]{2004Lazzarin}
{Lazzarin}, M., {Marchi}, S., {Barucci}, M.~A., {Di Martino}, M., \&
  {Barbieri}, C. 2004, \icarus, 169, 373, \dodoi{10.1016/j.icarus.2003.12.023}

\bibitem[{{Lazzarin} {et~al.}(2005){Lazzarin}, {Marchi}, {Magrin}, \&
  {Licandro}}]{2005Lazzarin}
{Lazzarin}, M., {Marchi}, S., {Magrin}, S., \& {Licandro}, J. 2005, \mnras,
  359, 1575, \dodoi{10.1111/j.1365-2966.2005.09006.x}

\bibitem[{{Lin} {et~al.}(2018){Lin}, {Ip}, {Lin}, {Cheng}, {Lin}, \&
  {Chang}}]{2018Lin}
{Lin}, C.-H., {Ip}, W.-H., {Lin}, Z.-Y., {et~al.} 2018, \planss, 152, 116,
  \dodoi{10.1016/j.pss.2017.12.019}

\bibitem[{{Lowry} {et~al.}(2007){Lowry}, {Fitzsimmons}, {Pravec},
  {Vokrouhlick{\'y}}, {Boehnhardt}, {Taylor}, {Margot}, {Gal{\'a}d}, {Irwin},
  {Irwin}, \& {Kusnir{\'a}k}}]{2007Lowry}
{Lowry}, S.~C., {Fitzsimmons}, A., {Pravec}, P., {et~al.} 2007, Science, 316,
  272, \dodoi{10.1126/science.1139040}

\bibitem[{{Lucas} {et~al.}(2019){Lucas}, {Emery}, {MacLennan},
  {Pinilla-Alonso}, {Cartwright}, {Lindsay}, {Reddy}, {Sanchez}, {Thomas}, \&
  {Lorenzi}}]{2018Lucas}
{Lucas}, M.~P., {Emery}, J.~P., {MacLennan}, E.~M., {et~al.} 2019, \icarus,
  322, 227, \dodoi{10.1016/j.icarus.2018.12.010}

\bibitem[{{Mainzer} {et~al.}(2019){Mainzer}, {Bauer}, {Cutri}, {Grav},
  {Kramer}, {Masiero}, {Sonnett}, \& {Wright}}]{2019Mainzer}
{Mainzer}, A., {Bauer}, J., {Cutri}, R., {et~al.} 2019, {NEOWISE Diameters and
  Albedos V2.0. urn:nasa:pds:neowise\_diameters\_albedos::2.0.}, NASA Planetary
  Data System

\bibitem[{{Masiero} {et~al.}(2017){Masiero}, {Nugent}, {Mainzer}, {Wright},
  {Bauer}, {Cutri}, {Grav}, {Kramer}, \& {Sonnett}}]{2017Masiero}
{Masiero}, J.~R., {Nugent}, C., {Mainzer}, A.~K., {et~al.} 2017, \aj, 154, 168,
  \dodoi{10.3847/1538-3881/aa89ec}

\bibitem[{{Masiero} {et~al.}(2018){Masiero}, {Redwing}, {Mainzer}, {Bauer},
  {Cutri}, {Grav}, {Kramer}, {Nugent}, {Sonnett}, \& {Wright}}]{2018Masiero}
{Masiero}, J.~R., {Redwing}, E., {Mainzer}, A.~K., {et~al.} 2018, \aj, 156, 60,
  \dodoi{10.3847/1538-3881/aacce4}

\bibitem[{{McNeill} {et~al.}(2019){McNeill}, {Hora}, {Gustafsson}, {Trilling},
  \& {Mommert}}]{2019McNeill}
{McNeill}, A., {Hora}, J.~L., {Gustafsson}, A., {Trilling}, D.~E., \&
  {Mommert}, M. 2019, \aj, 157, 164, \dodoi{10.3847/1538-3881/ab0e6e}

\bibitem[{{McNeill} {et~al.}(2018){McNeill}, {Fitzsimmons}, {Jedicke},
  {Lacerda}, {Lilly}, {Thompson}, {Trilling}, {DeMooij}, {Hooton}, \&
  {Watson}}]{2018McNeill}
{McNeill}, A., {Fitzsimmons}, A., {Jedicke}, R., {et~al.} 2018, \aj, 156, 282,
  \dodoi{10.3847/1538-3881/aaeb8c}

\bibitem[{{Mommert}(2017)}]{2017Mommert}
{Mommert}, M. 2017, {PHOTOMETRYPIPELINE: Automated photometry pipeline}.
\newblock \doeprint{1703.004}

\bibitem[{{Mueller} {et~al.}(2006){Mueller}, {Harris}, {Bus}, {Hora}, {Kassis},
  \& {Adams}}]{2006Mueller}
{Mueller}, M., {Harris}, A.~W., {Bus}, S.~J., {et~al.} 2006, \aap, 447, 1153,
  \dodoi{10.1051/0004-6361:20053742}

\bibitem[{{Nugent} {et~al.}(2016){Nugent}, {Mainzer}, {Bauer}, {Cutri},
  {Kramer}, {Grav}, {Masiero}, {Sonnett}, \& {Wright}}]{2016Nugent}
{Nugent}, C.~R., {Mainzer}, A., {Bauer}, J., {et~al.} 2016, \aj, 152, 63,
  \dodoi{10.3847/0004-6256/152/3/63}

\bibitem[{{Okamura} {et~al.}(2014{\natexlab{a}}){Okamura}, {Hasegawa}, {Usui},
  {Hiroi}, {Ootsubo}, {M{\"u}ller}, \& {Sugita}}]{2014bOkamura}
{Okamura}, N., {Hasegawa}, S., {Usui}, F., {et~al.} 2014{\natexlab{a}}, in
  Lunar and Planetary Science Conference, 1375

\bibitem[{{Okamura} {et~al.}(2014{\natexlab{b}}){Okamura}, {Sugita}, {Kamata},
  {Usui}, {Hiroi}, {Ootsubo}, {M{\"u}ller}, {Sakon}, \&
  {Hasegawa}}]{2014Okamura}
{Okamura}, N., {Sugita}, S., {Kamata}, S., {et~al.} 2014{\natexlab{b}}, in
  Lunar and Planetary Science Conference, 2446

\bibitem[{{Perna} {et~al.}(2018){Perna}, {Barucci}, {Fulchignoni}, {Popescu},
  {Belskaya}, {Fornasier}, {Doressoundiram}, {Lantz}, \& {Merlin}}]{2018Perna}
{Perna}, D., {Barucci}, M.~A., {Fulchignoni}, M., {et~al.} 2018, \planss, 157,
  82, \dodoi{10.1016/j.pss.2018.03.008}

\bibitem[{{Polishook}(2012)}]{2012Polishook}
{Polishook}, D. 2012, Minor Planet Bulletin, 39, 187

\bibitem[{{Pravec} {et~al.}(2012){Pravec}, {Harris}, {Ku{\v{s}}nir{\'a}k},
  {Gal{\'a}d}, \& {Hornoch}}]{2012Pravec}
{Pravec}, P., {Harris}, A.~W., {Ku{\v{s}}nir{\'a}k}, P., {Gal{\'a}d}, A., \&
  {Hornoch}, K. 2012, \icarus, 221, 365, \dodoi{10.1016/j.icarus.2012.07.026}

\bibitem[{{Pravec} {et~al.}(2005){Pravec}, {Harris}, {Scheirich}, {Ku{\v
  s}nir{\'a}k}, {{\v S}arounov{\'a}}, {Hergenrother}, {Mottola}, {Hicks},
  {Masi}, {Krugly}, {Shevchenko}, {Nolan}, {Howell}, {Kaasalainen},
  {Gal{\'a}d}, {Brown}, {DeGraff}, {Lambert}, {Cooney}, \&
  {Foglia}}]{2005Pravec}
{Pravec}, P., {Harris}, A.~W., {Scheirich}, P., {et~al.} 2005, \icarus, 173,
  108, \dodoi{10.1016/j.icarus.2004.07.021}

\bibitem[{{Pravec} {et~al.}(2006){Pravec}, {Scheirich}, {Ku{\v s}nir{\'a}k},
  {{\v S}arounov{\'a}}, {Mottola}, {Hahn}, {Brown}, {Esquerdo}, {Kaiser},
  {Krzeminski}, {Pray}, {Warner}, {Harris}, {Nolan}, {Howell}, {Benner},
  {Margot}, {Gal{\'a}d}, {Holliday}, {Hicks}, {Krugly}, {Tholen}, {Whiteley},
  {Marchis}, {DeGraff}, {Grauer}, {Larson}, {Velichko}, {Cooney}, {Stephens},
  {Zhu}, {Kirsch}, {Dyvig}, {Snyder}, {Reddy}, {Moore}, {Gajdo{\v s}},
  {Vil{\'a}gi}, {Masi}, {Higgins}, {Funkhouser}, {Knight}, {Slivan}, {Behrend},
  {Grenon}, {Burki}, {Roy}, {Demeautis}, {Matter}, {Waelchli}, {Revaz},
  {Klotz}, {Rieugn{\'e}}, {Thierry}, {Cotrez}, {Brunetto}, \&
  {Kober}}]{2006Pravec}
{Pravec}, P., {Scheirich}, P., {Ku{\v s}nir{\'a}k}, P., {et~al.} 2006, \icarus,
  181, 63, \dodoi{10.1016/j.icarus.2005.10.014}

\bibitem[{{Ryan} \& {Woodward}(2010)}]{2010Ryan}
{Ryan}, E.~L., \& {Woodward}, C.~E. 2010, \aj, 140, 933,
  \dodoi{10.1088/0004-6256/140/4/933}

\bibitem[{{Sanchez} {et~al.}(2013){Sanchez}, {Michelsen}, {Reddy}, \&
  {Nathues}}]{2013Sanchez}
{Sanchez}, J.~A., {Michelsen}, R., {Reddy}, V., \& {Nathues}, A. 2013, \icarus,
  225, 131, \dodoi{10.1016/j.icarus.2013.02.036}

\bibitem[{{Somers} {et~al.}(2010){Somers}, {Hicks}, {Lawrence}, {Rhoades},
  {Mayes}, {Barajas}, {McAuley}, {Foster}, {Shitanishi}, \&
  {Truong}}]{2010Somers}
{Somers}, J.~M., {Hicks}, M., {Lawrence}, K., {et~al.} 2010, in AAS/Division
  for Planetary Sciences Meeting Abstracts \#42, AAS/Division for Planetary
  Sciences Meeting Abstracts, 13.16

\bibitem[{{Stuart} \& {Binzel}(2004)}]{2004Stuart}
{Stuart}, J.~S., \& {Binzel}, R.~P. 2004, \icarus, 170, 295,
  \dodoi{10.1016/j.icarus.2004.03.018}

\bibitem[{{Tardioli} {et~al.}(2017){Tardioli}, {Farnocchia}, {Rozitis},
  {Cotto-Figueroa}, {Chesley}, {Statler}, \& {Vasile}}]{2017Tardioli}
{Tardioli}, C., {Farnocchia}, D., {Rozitis}, B., {et~al.} 2017, \aap, 608, A61,
  \dodoi{10.1051/0004-6361/201731338}

\bibitem[{{Thirouin} {et~al.}(2016){Thirouin}, {Moskovitz}, {Binzel},
  {Christensen}, {DeMeo}, {Person}, {Polishook}, {Thomas}, {Trilling}, \&
  {Willman}}]{2016Thirouin}
{Thirouin}, A., {Moskovitz}, N., {Binzel}, R.~P., {et~al.} 2016, \aj, 152, 163,
  \dodoi{10.3847/0004-6256/152/6/163}

\bibitem[{{Tholen}(1984)}]{1984Tholen}
{Tholen}, D.~J. 1984, PhD thesis, University of Arizona, Tucson

\bibitem[{{Thomas} {et~al.}(2014){Thomas}, {Emery}, {Trilling}, {Delb{\'o}},
  {Hora}, \& {Mueller}}]{2014Thomas}
{Thomas}, C.~A., {Emery}, J.~P., {Trilling}, D.~E., {et~al.} 2014, \icarus,
  228, 217, \dodoi{10.1016/j.icarus.2013.10.004}

\bibitem[{{Thomas} {et~al.}(2011){Thomas}, {Trilling}, {Emery}, {Mueller},
  {Hora}, {Benner}, {Bhattacharya}, {Bottke}, {Chesley}, \&
  {Delb{\'o}}}]{2011Thomas}
{Thomas}, C.~A., {Trilling}, D.~E., {Emery}, J.~P., {et~al.} 2011, \aj, 142,
  85, \dodoi{10.1088/0004-6256/142/3/85}

\bibitem[{{Trilling} {et~al.}(2010){Trilling}, {Mueller}, {Hora}, {Harris},
  {Bhattacharya}, {Bottke}, {Chesley}, {Delbo}, {Emery}, \&
  {Fazio}}]{2010Trilling}
{Trilling}, D.~E., {Mueller}, M., {Hora}, J.~L., {et~al.} 2010, \aj, 140, 770,
  \dodoi{10.1088/0004-6256/140/3/770}

\bibitem[{{Trilling} {et~al.}(2016){Trilling}, {Mommert}, {Hora}, {Chesley},
  {Emery}, {Fazio}, {Harris}, {Mueller}, \& {Smith}}]{2016Trilling}
{Trilling}, D.~E., {Mommert}, M., {Hora}, J., {et~al.} 2016, \aj, 152, 172,
  \dodoi{10.3847/0004-6256/152/6/172}

\bibitem[{{Vere{\v{s}}} {et~al.}(2015){Vere{\v{s}}}, {Jedicke}, {Fitzsimmons},
  {Denneau}, {Granvik}, {Bolin}, {Chastel}, {Wainscoat}, {Burgett}, \&
  {Chambers}}]{2015Veres}
{Vere{\v{s}}}, P., {Jedicke}, R., {Fitzsimmons}, A., {et~al.} 2015, \icarus,
  261, 34, \dodoi{10.1016/j.icarus.2015.08.007}

\bibitem[{{Warner}(2004)}]{2004Warner}
{Warner}, B.~D. 2004, Minor Planet Bulletin, 31, 67

\bibitem[{{Warner}(2015{\natexlab{a}})}]{2015cWarner}
---. 2015{\natexlab{a}}, Minor Planet Bulletin, 42, 41

\bibitem[{{Warner}(2015{\natexlab{b}})}]{2015hWarner}
---. 2015{\natexlab{b}}, Minor Planet Bulletin, 42, 115

\bibitem[{{Warner}(2016{\natexlab{a}})}]{2016cWarner}
---. 2016{\natexlab{a}}, Minor Planet Bulletin, 43, 66

\bibitem[{{Warner}(2016{\natexlab{b}})}]{2016hWarner}
---. 2016{\natexlab{b}}, Minor Planet Bulletin, 43, 143

\bibitem[{{Warner}(2017{\natexlab{a}})}]{2017bWarner}
---. 2017{\natexlab{a}}, Minor Planet Bulletin, 44, 335

\bibitem[{{Warner}(2017{\natexlab{b}})}]{2017cWarner}
---. 2017{\natexlab{b}}, Minor Planet Bulletin, 44, 22

\bibitem[{{Warner}(2017{\natexlab{c}})}]{2017iWarner}
---. 2017{\natexlab{c}}, Minor Planet Bulletin, 44, 98

\bibitem[{{Warner}(2018{\natexlab{a}})}]{2018Warner}
---. 2018{\natexlab{a}}, Minor Planet Bulletin, 45, 248

\bibitem[{{Warner}(2018{\natexlab{b}})}]{2018hWarner}
---. 2018{\natexlab{b}}, Minor Planet Bulletin, 45, 248

\bibitem[{{Warner} {et~al.}(2009){Warner}, {Harris}, \& {Pravec}}]{2009Warner}
{Warner}, B.~D., {Harris}, A.~W., \& {Pravec}, P. 2009, \icarus, 202, 134,
  \dodoi{10.1016/j.icarus.2009.02.003}

\bibitem[{{Waszczak} {et~al.}(2015){Waszczak}, {Chang}, {Ofek}, {Laher},
  {Masci}, {Levitan}, {Surace}, {Cheng}, {Ip}, {Kinoshita}, {Helou}, {Prince},
  \& {Kulkarni}}]{2015Waszczak}
{Waszczak}, A., {Chang}, C.-K., {Ofek}, E.~O., {et~al.} 2015, \aj, 150, 75,
  \dodoi{10.1088/0004-6256/150/3/75}

\bibitem[{{Weidenschilling} {et~al.}(1990){Weidenschilling}, {Chapman},
  {Davis}, {Greenberg}, {Levy}, {Binzel}, {Vail}, {Magee}, \&
  {Spaute}}]{1990Weidenschilling}
{Weidenschilling}, S.~J., {Chapman}, C.~R., {Davis}, D.~R., {et~al.} 1990,
  \icarus, 86, 402, \dodoi{10.1016/0019-1035(90)90227-Z}

\bibitem[{{Whiteley}(2001)}]{2001Whiteley}
{Whiteley}, Jr., R.~J. 2001, PhD thesis, University of Hawai'i at Manoa

\bibitem[{{Wisniewski} {et~al.}(1997){Wisniewski}, {Micha{\l}owski}, {Harris},
  \& {McMillan}}]{1997Wisniewski}
{Wisniewski}, W.~Z., {Micha{\l}owski}, T.~M., {Harris}, A.~W., \& {McMillan},
  R.~S. 1997, \icarus, 126, 395, \dodoi{10.1006/icar.1996.5665}

\bibitem[{{Ye} {et~al.}(2009){Ye}, {Shi}, {Xu}, {Lin}, \& {Zhang}}]{2009Ye}
{Ye}, Q., {Shi}, L., {Xu}, W., {Lin}, H.-C., \& {Zhang}, J. 2009, Minor Planet
  Bulletin, 36, 180

\bibitem[{{Zappala} {et~al.}(1990){Zappala}, {Cellino}, {Barucci},
  {Fulchignoni}, \& {Lupishko}}]{1990Zappala}
{Zappala}, V., {Cellino}, A., {Barucci}, A.~M., {Fulchignoni}, M., \&
  {Lupishko}, D.~F. 1990, \aap, 231, 548

\end{thebibliography}

%%%%%%%%%%%%%%%%%%%%%%%%%%%%%%%%%%%

\startlongtable
\begin{deluxetable*}{ccccccccc}
\tablecaption{DCT Observations Log \label{tab:dct_obs}}
\tablecolumns{10}
\tablenum{2}
\tablewidth{0pt}
\tablehead{
\colhead{\#} 	& \colhead{Designation} 		& \colhead{UTC}					& \colhead{V}					&\colhead{$\alpha$}					& \colhead{Airmass}		& \colhead{Exptime} 			& \colhead{Duration}  & \colhead{$\Delta$m}\\
\colhead{}		& \colhead{}		 			& \colhead{\footnotesize{(yymmdd)}}		& \colhead{\footnotesize{(mag)}}	& \colhead{\footnotesize{($^{\circ}$)}}	& \colhead{}			& \colhead{\footnotesize{(s)}}	& \colhead{\footnotesize{(hour)}}	& \colhead{\footnotesize{(mag)}}
}	
\startdata
3199 	&1982 RA			& 161129		& 18.6		&32.7	& 1.4		& 60		&0.33	&0.13	\\
3691 	&1982 FT			& 161129		& 17.8		&15.2	& 1.2		& 60		&0.30	&0.31	\\
5131 	&1990 BG			& 161129		& 18.0		&18.5	& 1.3		& 60		&0.33	&0.82	\\
5604 	&1992 FE			& 170403	 	& 16.4		&41.6	& 1.2		& 60		&0.22	&0.11	 \\
\vdots	& \vdots				 &  170403 	& 16.4		&41.6	& 1.5		& 60		&0.21	&0.08	\\
5751 	&1992 AC			& 170629  	& 18.7		&10.0	& 1.6		& 60		&0.45	&0.11	 \\
10165 	&1995 BL2			&  170403		& 19.5		&41.0	& 1.3		& 60		&0.21	&0.23	\\
23183 	&2000 OY21		 	& 170325 		& 21.1		&22.8	& 1.5		& 60		&0.24	&0.97	\\
\vdots	& \vdots			 	&  170517 	& 20.2		&22.7	& 1.1		& 60		&0.54	&1.03	\\
\vdots	& \vdots			 	& 170629 		& 19.1		&11.7	& 1.3		& 60		&0.19	&0.76	\\
24443	& 2000 OG		 	& 170517 		& 21.9		&9.3		& 1.2		& 60		&0.52	&0.08	\\
54789	& 2001 MZ7		 	&  170403		& 17.3		&17.3	& 1.2		& 30		&0.13	&0.06	\\
\vdots	& \vdots			 	&  170403		& 17.3		&17.3	& 1.5		& 30		&0.63	&0.14	\\
\vdots	& \vdots			 	& 170517 		& 17.7		&47.6	& 1.2		& 30		&0.14	&0.64	\\
87024	& 2000 JS66		 	&  170403		& 19.0		&35.6	& 2.2		& 45		&0.63	&0.84	\\
\vdots	& \vdots			 	& 170517 		& 20.1		&55.5	& 1.8		& 30		&0.37	&0.86	\\
90373	& 2003 SZ219	 		& 170325		& 21.7		&26.0	& 1.5 	& 60		&0.54	&0.42	\\
\vdots	& \vdots			 	&  170403		& 21.5		&23.1	& 1.6		& 60		&0.21	&0.47	\\
\vdots	& \vdots			 	& 170517 		& 20.5		&20.5	& 1.5		& 60		&0.33	&0.18	\\
136818 	&1997 MW1 			& 170629 		& 17.9		&27.6	& 1.2		& 15		& 0.58	&0.80	\\
162882 	&2001 FD58	  		&  170403		& 19.3		&7.1		& 1.4		& 30		&0.21	&0.07	\\
162980 	&2001 RR17 	  		& 150920		& 19.7		&32.0	& 1.3		& 60		&0.09	&0.48	\\
185716 	&1998 SF35	  		& 170325		& 21.1		&32.8	& 1.6		& 60		&0.60	&0.42	\\
\vdots	& \vdots				 &  170403	& 21.0		&31.1	& 1.6		& 60		&0.34	&0.46	\\ 
\vdots	& \vdots				 & 170517 	& 20.4		&19.2	& 1.2		& 45		&0.21	&0.34	 \\
\vdots	& \vdots				 & 170629 	& 21.0		&23.7	& 1.0		& 60		&0.21	&0.45	\\
222073 	&1999 HY1		 	& 170517 		& 17.8		&31.6	& 1.2		& 15		&0.87	&0.44	\\
\vdots	& \vdots				 & 170629 	& 19.5		&54.7	& 1.0		& 15		&1.52	&0.62	\\
226198 	&2002 UN3		   	&  170403		& 19.6		&40.0	& 1.0		& 45		&0.27	&0.24	\\
232368 	&2003 AZ2		  	& 170325		& 20.5		&44.2	& 1.3		& 45		&0.28	&0.36	 \\
\vdots	& \vdots				 &  170403	& 20.3		&45.3	& 1.3		& 30		&0.32	&0.60	\\
\vdots	& \vdots				 & 170629 	& 18.0		&58.7	& 1.0		& 30		&1.52	&0.36	\\
285638 	&2000 SO10	 		& 150920		& 20.2		&19.3	& 1.0		& 30		&1.90	&0.25	\\
311066 	&2004 DC		 	& 150920		& 22.4		&25.5	& 1.1		& 60		&0.09	&0.49	 \\
354876 	&2006 BG55	 		& 170403		& 18.6		&71.0	& 1.6		& 30		&0.13	&0.54	\\
416151 	&2002 RQ25	 		& 150920		& 22.3 		&44.8	& 1.0		& 60		&0.09	&0.81	
\enddata
\tablecomments{UTC date of observations and V-magnitude from Horizons. Airmass from midpoint of observations. Exposure time of individual frames. Duration is net integration time on object and $\Delta m$ is the $max-min$ lower limit amplitude from the observations.}
\end{deluxetable*}

\startlongtable
\begin{deluxetable*}{ccccccccc}
\tablecaption{Calar Alto Observations Log \label{tab:caloalto_obs}}
\tablecolumns{9}
\tablenum{3}
\tablewidth{0pt}
\tablehead{
\colhead{\#}	&\colhead{Designation} 	& \colhead{UTC}		& \colhead{V}					&\colhead{$\alpha$}					& \colhead{Airmass}	& \colhead{Exptime} 				& \colhead{Duration} 			& \colhead{$\Delta$m} \\
\colhead{}		&\colhead{}		 	& \colhead{yymmdd}		& \colhead{\footnotesize{(mag)}} 	& \colhead{\footnotesize{($^{\circ}$)}}	& \colhead{}		& \colhead{\footnotesize{(s)}}		& \colhead{\footnotesize{(hour)}} 	& \colhead{\footnotesize{(mag)}}
}	
\startdata
3691 		&1982 FT		& 160728		& 19.6	 	&23.6	&1.4		&300		& 0.70		&0.07 		\\
\vdots 		& \vdots 	& 160729		& 19.6		& 23.7	&1.5		&180 	&1.17		&0.46 		\\
\vdots 		& \vdots 	& 160730		& 19.6		& 23.8	&2.0		& 180	&1.30		&0.20		\\
\vdots 		& \vdots 	& 160801		& 19.6		&24.0	&1.7		& 180	&1.71		& 0.17		\\
\vdots 		& \vdots 	& 160804		& 19.6		&24.4	&1.5		& 180	&1.64		&0.45 		\\
\vdots 		& \vdots 	& 160805		& 19.6		&24.5	&1.6		&180 	&2.19		&0.37 		\\
\vdots 		& \vdots 	& 160806		& 19.6		&24.6	&1.7		& 180	&1.72		&0.13 		\\
\vdots 		& \vdots 	& 160807		& 19.6		&24.7	&1.5		&180 	&2.72		&0.50 		\\
\vdots 		& \vdots 	& 160808		& 19.6		&24.8	&1.7		& 180	&2.78		&0.34 		\\
\vdots 		& \vdots 	& 161226		& 17.9		&19.1	&1.0		& 150	&9.0			&0.11 		\\
\vdots 		& \vdots 	& 161227		& 17.9		&19.4	&1.0		& 150	&7.63		&0.12 		\\
\vdots 		& \vdots 	& 161229		& 18.0		&20.0	&1.1		& 150	&7.17		&0.30 		\\
\vdots 		& \vdots 	& 161230		& 18.0		&20.2	&1.1		& 150	&6.64		& 0.16		\\
5751 		&1992 AC	 	& 151203  	& 18.7		&25.1	&1.8		& 300	&3.90		&0.14		 \\
\vdots 		& \vdots 	& 151206  	& 18.8		&25.6	&1.8		& 300	&3.79		&0.21 		 \\
\vdots 		& \vdots 	& 151208  	& 18.8		&25.9	&1.8		& 300	&2.62		&0.08 		 \\
\vdots 		& \vdots 	& 151215  	& 18.9		&26.9	&1.7		& 300 	&2.53		&0.11 		 \\
\vdots 		& \vdots 	& 151216  	& 18.9		&27.0	&1.7		& 300	&2.33		&0.19		 \\
**137052 		&1998 VO33	& 151107 		& 20.1		&26.0	&1.1		& 300	&5.19		&0.25		\\
\vdots 		& \vdots 	& 151108 		& 20.1		&25.6	&1.1		& 300	&5.91		& 0.46		\\
\vdots 		& \vdots 	& 151118 		& 20.5		&20.5	&1.0		& 300	&4.78		&0.49 		\\
\vdots 		& \vdots 	& 151206 		& 19.5		&9.2		&1.0		& 300	&9.20		&0.37 		\\
\vdots 		& \vdots 	& 151210 		& 19.4		&6.8		&1.0		& 300	&1.97		&0.64 		\\
\vdots 		& \vdots 	& 151213 		& 19.3		&5.3		&1.0		& 300	&5.07		&1.15 		\\
\vdots 		& \vdots 	& 151214 		& 19.3		&4.9		&1.0		& 300	&9.10		&0.35 		\\
\vdots 		& \vdots 	& 151217 		& 19.3		&4.4		&1.2		& 300	&8.75		& 0.42		\\
162980 		&2001 RR17	& 160408		& 18.2		&75.5	&1.5		& 300	&1.81		& 0.16		\\
\vdots 		& \vdots 	& 160409		& 18.2		&74.5	&1.5		& 300	&1.83		& 0.22		\\
\vdots 		& \vdots 	& 160410		& 18.2		&73.8	&1.6		& 300	&1.07		& 0.45		\\
\vdots 		& \vdots 	& 160411		& 18.2		&73.0	&1.5		& 300	&1.72		& 0.29		\\
\vdots 		& \vdots 	& 160413		& 18.3		&71.5	&1.3		& 300	&2.15		& 0.40		\\
\vdots 		& \vdots 	& 160414		& 18.3		&70.8	&1.5		& 300	&1.64		& 0.19		\\
\vdots 		& \vdots 	& 160415		& 18.3		&70.0	&1.7		& 300	&0.47		& 0.31		\\
\vdots 		& \vdots 	& 160430		& 18.5		&61.1	&1.5		& 300	&2.71		& 0.28		\\
\vdots 		& \vdots 	& 160502		& 18.6		&60.1	&1.4		& 300	&0.70		& 0.19		\\
\vdots 		& \vdots 	& 160728		& 19.0		&27.5	&1.8		& 130	&0.91		& 0.30		\\
\vdots 		& \vdots 	& 160731		& 19.0		&26.9	&1.6		& 130	&0.91		&0.33		\\
\vdots 		& \vdots 	& 160801		& 19.0		&26.7	&1.6		& 130	&1.35		& 0.65		\\
\vdots 		& \vdots 	& 160804		& 19.1		&26.2	&1.6		& 130	&1.29		& 0.51		\\
\vdots 		& \vdots 	& 160806		& 19.1		&25.9	&1.6		& 130	&1.29		& 0.51		\\
\vdots 		& \vdots 	& 160807		& 19.1		&25.8	&1.6		& 130	&1.33		&0.50		\\
275792 		&1998 SF35 	& 160728		& 21.0		&23.7	&1.3		& 100	&0.70		& 0.15		\\
\vdots 		& \vdots 	& 160729		& 21.0		&23.8	&1.3		& 300	&0.89		& 0.17		\\
\vdots 		& \vdots 	& 160730		& 21.0		&24.0	&1.3		& 100	&0.07		& 0.14		\\
\vdots 		& \vdots 	& 160804		& 21.0		&24.7	&1.2		& 300	&2.35		& 0.26		\\
\vdots 		& \vdots 	& 160805		& 21.0		&24.8	&1.2		& 300	&2.26		& 0.38		\\
**285638 		&2000 SO10	& 151015		& 20.0		&15.2	&1.0		& 300	&3.82		& 0.63		\\
\vdots 		& \vdots 	& 151109		& 20.5		&23.4	&1.0		& 300	&6.17		& 1.08		\\
\vdots 		& \vdots 	& 151110		& 20.5		&23.8	&1.3		& 300	&1.36		& 0.84		\\
\vdots 		& \vdots 	& 151111		& 20.5		&24.1	&1.0		& 300	&4.80		& 0.81		\\
\vdots 		& \vdots 	& 151113		& 20.5		&24.8	&1.0		& 300	&5.67		& 0.95		\\
\vdots 		& \vdots 	& 151114		& 20.6		&25.1	&1.0		& 300	&5.20		& 1.06		\\
\vdots 		& \vdots 	& 151115		& 20.6		&25.4	&1.0		& 300	&4.61		&1.16 		\\
354876 		&2006 BG55	& 151208		& 19.3		&18.9	&1.4		& 300 	&5.12		& 0.67		\\
\vdots 		& \vdots 	& 151210		& 19.4		&17.7	&1.3		& 300 	&1.39		& 0.13		\\
\vdots 		& \vdots 	& 151211		& 19.4		&17.2	&1.4		& 300	& 2.09		& 0.15		\\
\vdots 		& \vdots 	& 151212		& 19.4		&16.7	&1.3		& 300	& 0.71		& 0.11		\\
\vdots 		& \vdots 	& 151213		& 19.4		&16.2	&1.3		& 300	& 3.84		& 0.23		\\
\vdots 		& \vdots 	& 151214		& 19.4		&15.9	&1.3		& 300	& 4.81		& 0.62		\\
\vdots 		& \vdots 	& 151215		& 19.4		&15.6	&1.4		& 300	& 4.03		& 0.22		\\
\vdots 		& \vdots 	& 151217		& 19.5		&15.2	&1.3		& 300	& 4.15		& 0.53		
\enddata
\tablecomments{UTC date of observations and V-magnitude from Horizons. Airmass from midpoint of observations. Exposure time of individual frames. Duration is net integration time on object and $\Delta m$ is the $max-min$ lower limit amplitude from the observations. ** Indicate the two objects which have full lightcurve coverage. 137052 has a derived period of $9.007\pm0.008$ hours \citep{2018Warner}. We derive in this work a preliminary period of $5.51\pm0.1$ hours for 285638. }
\end{deluxetable*}

\startlongtable
\begin{deluxetable*}{ccccc}
\tablecaption{LCDB Lower Limit Amplitudes \label{tab:LCDB}}
\tablecolumns{5}
\tablenum{4}
\tablewidth{0pt}
\tablehead{
\colhead{\#}	& \colhead{Designation} 	& \colhead{$\alpha$}						& \colhead{Amplitude}			& \colhead{Photometry Reference} \\ 
\colhead{	}	& \colhead{}		 	& \colhead{\footnotesize{($^{\circ}$)}}		& \colhead{\footnotesize{(mag)}}	& \colhead{}
}	
\startdata
1620			&1951 RA			&19.0				&1.0			& [19] \\
1865			&1971 UA			&21.8				&2.1 			& [20] \\
1915			&1953 EA			&33.3				&0.2			& [2] \\
3671			&1984 KD			&37.1				&0.23		& [7]\\
5879			&1992 CH1			&23.2				&0.45		& [14]\\
17511		&1992 QN			&11					&0.84		& [18]\\
25143		&1998 SF36			&5.0					&0.70		& [9]\\
33342		&1998 WT24			&69.8 				&0.29		& [13]\\
42286		&2001 TN41			&14.8				&0.44		& [16]\\
53110		&1999 AR7			&32.7				&0.37		& [10]\\
54509		&2000 PH5			&31.9				&0.83		& [4]\\
136849		&1998 CS1			&22.8				&0.16		& [21]\\
138893		&2000 YH66			&32.1				&0.7			& [5] \\
162004		&1991 VE			&41.8 				&1.11 		& [11] \\
172034		&2001 WR1			&25.3				&0.95		& [17]\\
257838		&2000 JQ66			&37.6 				&0.63 		&[15]\\
374158		&2004 UL			&21.2				&1.2			& [11]\\
399167		&2015 FG36			&58.1				&0.59		& [8]\\
416186		&2002 TD60			&3					&1.4			& [6]\\
441825		&2009 SK1			&8.7					&0.6 			&[12]\\
443880		&2001 UZ16			&84.5				&0.97		&[12]\\
			&2001 SQ3			&35.7				&0.39 		&[3]\\
			&2010 NR1			&20.8				&1.8 			&[1]\\
			&2014 YD			&72.3				&0.2 			&[8]\\
			&2015 FS332			&64.5				&0.84		& [12]\\
\enddata
\tablecomments{Lightcurve amplitudes and phase angle are from the Asteroid Lightcurve Database \citep{2009Warner}.}
\tablereferences{	[1]~\cite{2010Farmer},
				[2]~\cite{1989Harris},
				[3]~\cite{2011Hergenrother},
				[4]~\cite{2007Lowry},
				[5]~\cite{2012Polishook},
				[6]~\cite{2005Pravec},
				[7]~\cite{2006Pravec},
				[8]~\cite{2016Thirouin},
				[9]~\cite{2004Warner},
				[10]~\cite{2015cWarner},
				[11]~\cite{2015hWarner},
				[12]~\cite{2016cWarner},
				[13]~\cite{2016hWarner},
				[14]~\cite{2017bWarner},
				[15]~\cite{2017cWarner},
				[16]~\cite{2017iWarner},
				[17]~\cite{2018hWarner},
				[18]~\cite{2015Waszczak},
				[19]~\cite{1990Weidenschilling},
				[20]~\cite{1997Wisniewski},
				[21]~\cite{2009Ye}
				}
\end{deluxetable*}

\end{document}